\documentclass[iop, apj, tighten, numberedappendix]{emulateapj}
\slugcomment{{To appear in the Astrophysical Journal.}}

\usepackage{amsmath,amsthm}
\usepackage{amssymb}

\usepackage{mathtools}
\usepackage{empheq}
\usepackage{bbm}
\usepackage{bm}

\usepackage{hyperref}

\usepackage{epstopdf}

\usepackage{color} 

\newcommand{\myemail}{vboening@leibniz-kis.de}

\newcommand\ii{{\rm i}}
\newcommand\id{{\mathrm d}}

\newcommand\cO{{\mathcal O}}

\newcommand\br{{\mathbf r}}

\newcommand\Power{{\mathcal{P}}}

\newcommand\bnabla{{\bm \nabla }}
\newcommand\bcdot{{\, \bm \cdot \,}}
\renewcommand\bv{{\bf v}}

\newcommand\bxi{{\bm \xi}}

\newcommand\bK{{ \bf K}}

\newcommand\Cref{C_{\text{ref}}}
\newcommand\cref{\Cref}

\newcommand\taudiff{\tau_{\text{diff}}}

\newcommand\Wdiffstar{{W^*_\text{diff}}}

\newcommand\bZ{{\bf Z}}

\newcommand\robs{{r_{\text{obs}}}}
\newcommand\EE{{\mathbb{E}}}

\renewcommand\apjl{ApJL}
\renewcommand\solphys{SoPh}

\shorttitle{Validation of Spherical Born Kernels}

\shortauthors{B\"oning et al.}

\begin{document}


\title{Validation of Spherical Born Approximation Sensitivity Functions for Measuring Deep Solar Meridional Flow}

\author{Vincent~G.~A.~B\"oning}
\affil{Kiepenheuer-Institut f\"ur Sonnenphysik, 79104 Freiburg, Germany}
\email{\myemail}
\author{Markus~Roth}
\affil{Kiepenheuer-Institut f\"ur Sonnenphysik, 79104 Freiburg, Germany}
\author{Jason Jackiewicz}
\affil{New Mexico State University, Las Cruces, NM 88001, USA}
\and
\author{Shukur Kholikov}
\affil{National Solar Observatory, Tucson, AZ 85719, USA}


\begin{abstract}

Accurate measurements of deep solar meridional flow are of vital interest for understanding the solar dynamo. 
In this paper, we validate a recently developed method for obtaining sensitivity functions (kernels) for travel-time measurements to solar interior flows using the Born approximation in spherical geometry
, which is expected to be more accurate than the classical ray approximation. 
Furthermore, we develop a numerical approach to efficiently compute a large number of kernels based on the separability of the eigenfunctions into their horizontal and radial dependence.
%
The validation is performed using a hydrodynamic simulation of linear wave propagation in the Sun, 
which includes a standard single-cell meridional flow profile.
%
We show that, using the Born approximation, it is possible to accurately model observational quantities relevant for time-distance helioseismology such as the mean power spectrum, disc-averaged cross-covariance functions, and travel times in the presence of a flow field. 
In order to closely match the model to observations, we show that it is beneficial to use mode frequencies and damping rates which were extracted from the measured power spectrum. 
Furthermore, the contribution of the radial flow to the total travel time is found to reach 20\% of the contribution of the horizontal flow at travel distances over $40\degr$. 
Using the Born kernels and a 2D SOLA inversion of travel times, we can recover most features of the input meridional flow profile.
%
The Born approximation is thus a promising method for inferring large-scale solar interior flows.

\end{abstract}




\section{INTRODUCTION}
\label{secintro}

Accurate inferences of solar meridional flow are important for modelling the solar dynamo (e.g., \citealp{Charbonneau2014}, \citealp{Charbonneau2010}). At or near the surface, the magnitude of the meridional flow is known to a good extent (e.g., \citealp{GB2005}, \citealp{Miesch2005LRSP}) from a variety of methods such as Doppler-shifts and feature tracking (e.g., \citealp{Ulrich2010,Hathaway2012}), time-distance helioseismology (e.g., \citealp{Giles1997Nature,Beck2002,Zhao2004a},) ring-diagram analysis (e.g., \citealp{Haber2002,Komm2005,GonzalezHernandez2008,GonzalezHernandez2010,Basu2010}), and Fourier-Hankel analysis \citep{Braun1998,Krieger2007}.

 The nature of meridional flows in deeper layers below about 0.9 solar radii still remains under debate. Such measurements were obtained using tracking of supergranules \citep{Hathaway2012} and helioseismologic techniques, such as time-distance helioseismology (\citealp{Giles1997Nature}; \citealp{Zhao2013,Jackiewicz2015,Rajaguru2015}), Fourier-Hankel analysis \citep{Braun1998}, and global helioseismology \citep{Schad2013}. The conclusions on the global pattern of the meridional flow are very different, ranging from multiple cells in depth \citep{Zhao2013} and multiple cells in depth and latitude \citep{Schad2013} to a single-cell picture with a return flow starting in rather shallow layers (\citealp{Jackiewicz2015}; \citealp{Hathaway2012}, at about 0.9 solar radii) or in deeper regions \citep[below 0.85 solar radii]{Giles1997Nature,Braun1998,Rajaguru2015}. However, there are several details to keep in mind concerning these inversion results \citep{Jackiewicz2015}, e.g., that uncertainties in the results due to systematic effects are likely to be larger than the random errors in the inversion results, that the results were obtained using different instruments, and that they cover different periods in time. Furthermore, the impact of the radial flow component on the travel times was not taken into account in the results obtained by \citet{Zhao2013} and \citet{Jackiewicz2015}.

In time-distance helioseismology \citep{Duvall1993}, flows can be inferred using sensitivity functions (kernels), which are a model for the impact of the flows on travel-time measurements of acoustic waves. 
Measurements of deep meridional flow have so far been done using the rather classical ray approximation (\citealp{Kosovichev1996}; \citealp{KosovichevDuvall1997}). In this model, travel times are assumed to be sensitive to flows only along an infinitely thin ray path which connects the two observation points.

Recently, \citet{Boening2016} extended a Born approximation model for the travel-time measurements from Cartesian to spherical geometry. An alternative approach for computing Born kernels was proposed very recently by \citet{Gizon2016}. These developments permit the use of Born approximation kernels for inferring the deep meridional flow. In the Born approximation (e.g., \citealp{Birch2000}, \citealp{GB2002}), the full wave field in the solar interior is modelled using a damped wave equation which is stochastically excited by convection. This wave equation is solved in zero-order and in its first-order perturbation, which includes advection in the presence of a flow field. When modelling travel times using the Born approximation, the advection and first-order scattering of the wave field at any location inside the Sun is thereby taken into account.

The ray approximation is expected to be accurate if the underlying flow field does not vary at length scales which are smaller than a wavelength \citep[e.g.,][]{Birch2001}. In the case of flows at the bottom of the convection zone, this scale was estimated to be of the order of $200\,\rm{Mm}$ (\citealp{Boening2016} and references therein). 
%
%
%
%
If the flow varies on smaller length scales, the Born approximation is thought to be more accurate (e.g., \citealp{Bogdan1997}, \citealp{Birch2004}, \citealp{Couvidat2006}, and \citealp{BG2007}).

In addition to modelling the perturbation to the full wave field in the solar interior, an advantage of the Born approximation is that it also provides a model for additional observational quantities, such as disc-averaged cross-covariances and mean power spectra, which is not the case for the ray approximation. The accuracy of the model can therefore easily be validated.

Several methods for inferring the deep solar meridional flow \citep{Zhao2013,Jackiewicz2015, Roth2016} have been validated using a linear numerical simulation of the solar interior wave field by \citet{Hartlep2013}, which includes a standard single-cell meridional flow profile. The same simulation is to be used in this paper in order to validate the use of spherical Born kernels for inferring solar meridional flows.

In Section~\ref{secborn}, we will give a short introduction to the Born approximation as used in time-distance helioseismology. Section~\ref{secfast} includes the details of a numerical optimization, which was necessary in order to obtain the required number of kernels in an acceptable amount of time. The main comparison of the Born approximation model with artificial data obtained from the simulation is presented in Section~\ref{secfwdcomp}. In Section~\ref{secradial}, we discuss the relevance of the radial flow component on the measured travel times. Furthermore, we show inversion results of this data obtained with the Born kernels in Section~\ref{secinversions}. Conclusions are presented in Section~\ref{secconclusions}.

\begin{figure*}%
\begin{center}

\includegraphics[width=0.33\textwidth]{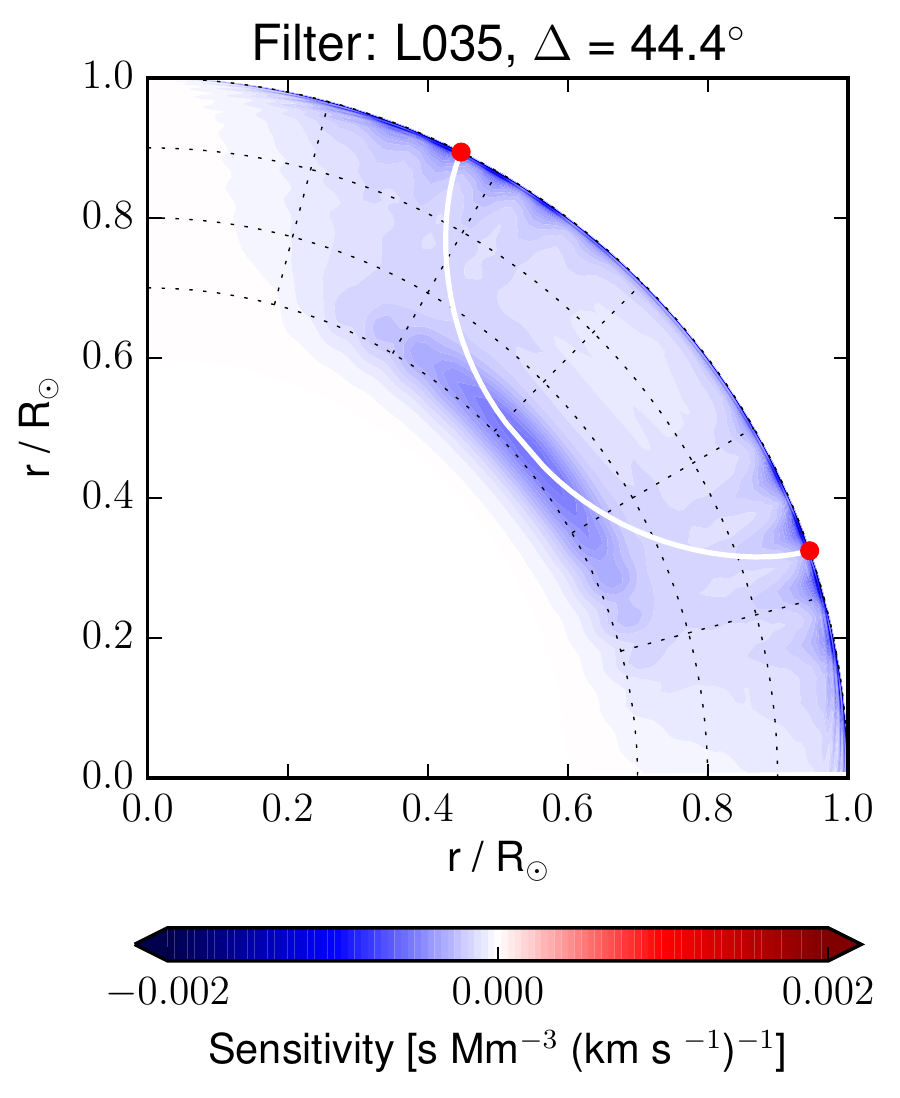}%
\includegraphics[width=0.33\textwidth]{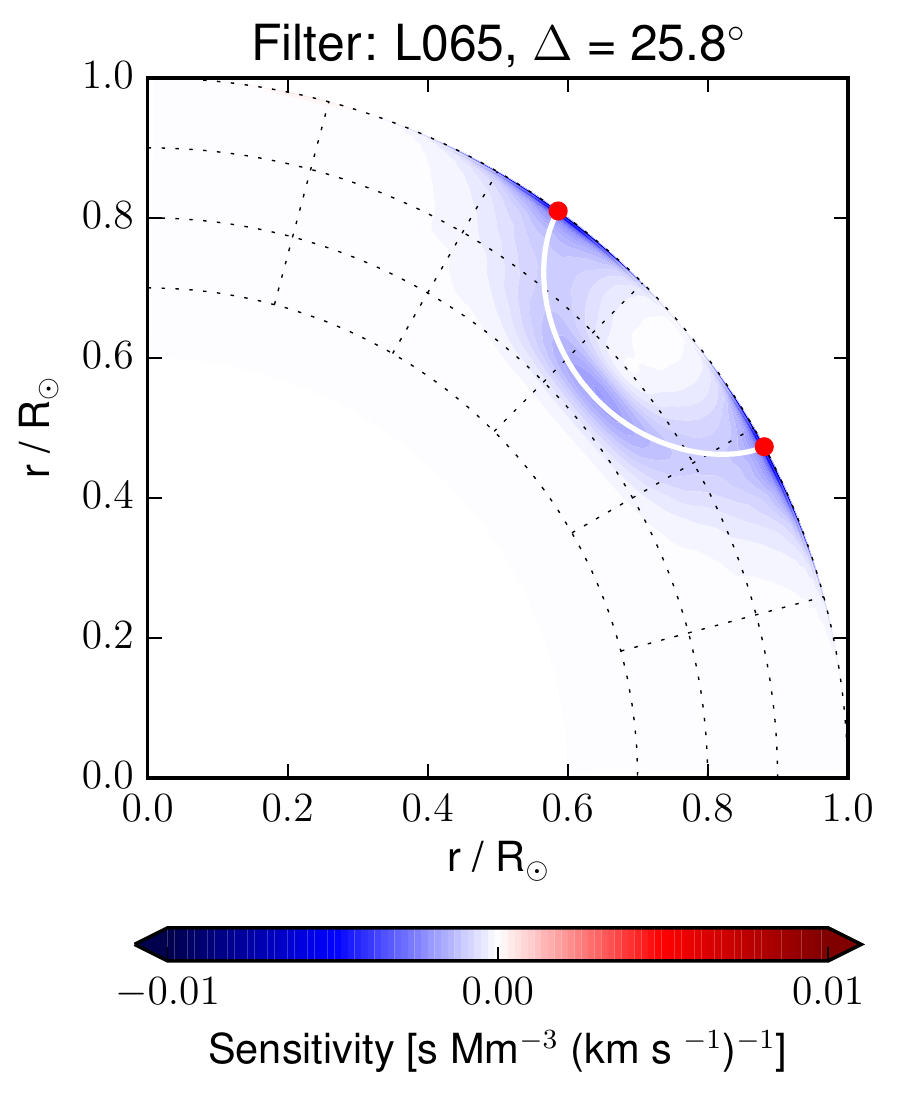}%
\includegraphics[width=0.33\textwidth]{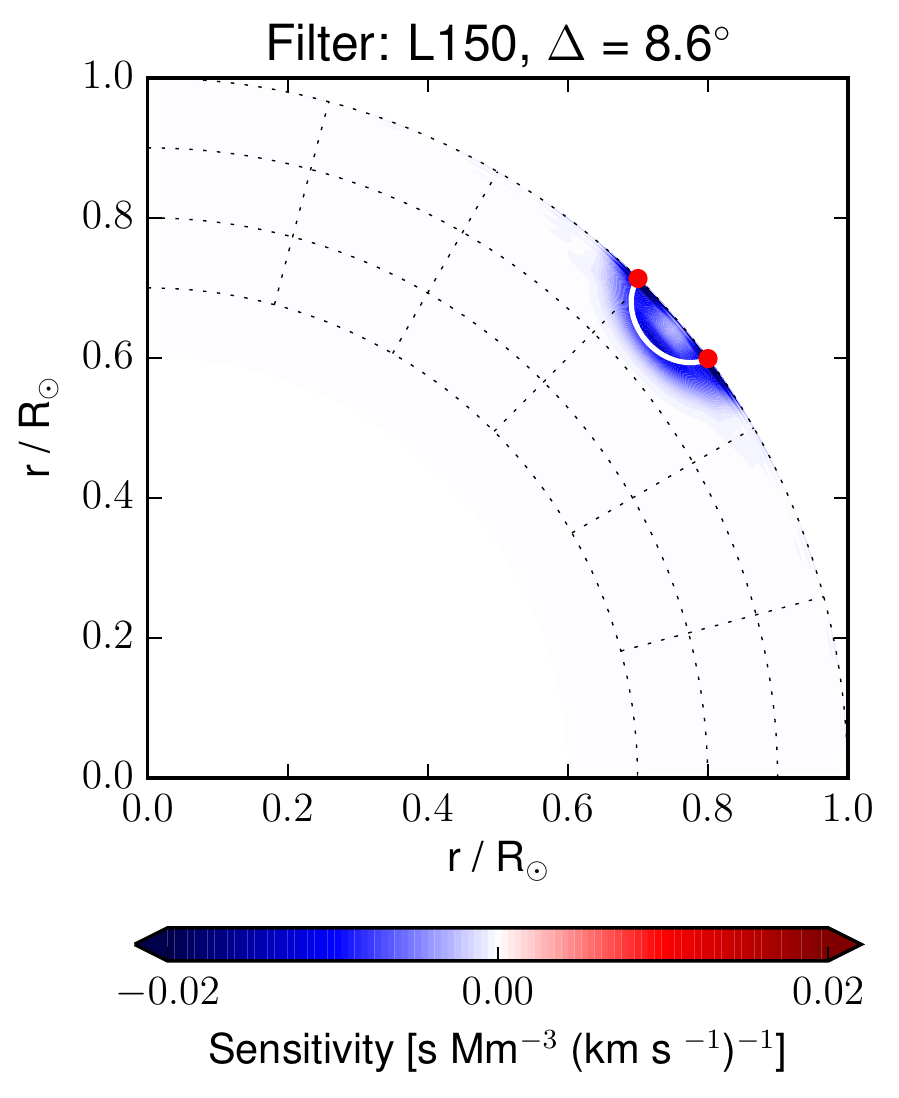}%

\end{center}

\caption{Example kernels for point-to-arc travel times for a central latitude of $40.5\degr$ and a travel distance given in each panel. The kernels were averaged in the same way as it is done for the travel-times in the data analysis procedure in Section~\ref{secfwdcomp} and integrated over longitude. A ray path connecting the nominal location of the observation points (red dots) is also shown (white line). See Table~\ref{tabfilters} for details on the filters used.\label{figkernels}}%
\end{figure*}

\section{THE BORN APPROXIMATION IN TIME-DISTANCE HELIOSEISMOLOGY}
\label{secborn}

We first briefly summarize the Born approximation model introduced by \citet{Boening2016}, which is to be validated in the following sections. Following \citet{GB2002}, the measurement process in time-distance helioseismology is modelled as closely to observations as possible. The objective of the model is to give a linear relationship between a small-magnitude flow field $\bv(\br)$ in the solar interior, which perturbs a mean observed or background model travel time, $\tau$, according to
\begin{equation}
\EE[\delta \tau] = \int_\Sun \bK(\br) \bcdot \bv(\br) \, \id^3 \br,
\label{eqkernelgoal}
\end{equation}
where $\EE$ denotes the expectation value of a stochastic quantity. Vector quantities are printed in bold throughout this work. Once an expression for the sensitivity function $\bK(\br)$ is found and travel times are measured, Equation~\eqref{eqkernelgoal} can be used to infer solar interior flows.


In order to achieve this goal, a model is first developed for an unperturbed spherically symmetric non-rotating Sun (zero-order problem). The stochastically driven and damped wave equation for solar oscillations is solved by expansion into solar eigenmodes. The resulting wave field $\bxi(\br,t)$ at location $\br=(r,\theta,\phi)=(r,\Omega)$ and time $t$ can be used to model the observed Doppler signal $\tilde\Phi$ via $\tilde\Phi=\dot \xi_r$, where the line-of-sight is assumed to be radial. After a spherical harmonic transform
\begin{equation}
\tilde a_{lm}(t) = \int_{S^2} Y_{lm}^*(\Omega) \tilde \Phi(\robs,\Omega,t) \, \id \Omega \label{eqSHT}
\end{equation}
and a Fourier transform (indicating the Fourier transform by the use of the variable $\omega$ instead of $t$), a power spectrum can be computed both from observations and from the model,
\begin{align}
   \Power(l,\omega) 
	= \;\frac{2\pi}{T} \sum_{m=-l}^l | a_{lm}(\omega)|^2, 
	\label{eqpowerfiltered}
\end{align}
where the filtered $a_{lm}(\omega)$ were obtained by multiplying $\tilde a_{lm}(\omega)$ by a filter function $f(l,\omega)$. After reconstructing the filtered Dopplergram time series, $\Phi(\theta,\phi,t)$,
\begin{align}
        \Phi(\theta,\phi,t) &= \sum_{l,m} a_{lm}(t) Y_{lm}(\phi,\theta),
        \label{eqdatafiltered2}
\end{align}
cross-covariances are computed by
\begin{equation}
        C(\br_1,\br_2,\omega) = \frac{2 \pi}{T} \Phi^*(\br_1,\omega)  \Phi(\br_2,\omega), \label{eqconvcorrw}
\end{equation}
which are used for fitting travel-times from observations, see \cite{GB2002}, \citet[hereafter GB04]{GB2004}, and \citet{Boening2016}. As our goal is to measure flows, we use travel-time differences $\taudiff=\tau_+ - \tau_-$, and the terms travel-time and travel-time difference are used interchangeably in this work.


In order to model the effect of a flow field on the travel times, the zero-order model for the mean Sun is perturbed. The flow is assumed to be small and thus only linear effects are taken into account. The perturbation of the flow is introduced in the wave equation by adding an advection term. This first-order wave equation is solved for the perturbation to the wave field, $\delta \bxi(\br,t)$. Taking only first-order terms (i.e. linear terms) into account, this corresponds to modelling the first-order scattering of the modes in the solar interior due to the flow field, see \cite{GB2002}. The perturbation to the wave field is then used to obtain the perturbation to the cross-covariance and the travel-time shift according to Equation~\eqref{eqkernelgoal}. The result is
\begin{align}
        \bK(\br_1,\br_2; \br)=& \sum_{j=(ln), i=(\bar l \bar n)} \; J_{ij}(\br_1,\br_2) \, \bZ^{ij}(\br_1,\br_2; \br)  \nonumber \\
				& \quad \quad + \Big( 1 \leftrightarrow 2 \Big)^* \label{eqcalckernel},
\end{align}
where the last term is identical to the previous term, apart from complex conjugation and exchange of indices 1 and 2. In Equation~\eqref{eqcalckernel}, the sum is taken over all pairs of eigenmodes $(j,i)$ and the quantities $J_{ij}$ and $\bZ^{ij}$ describe the scattering of mode $j$ into mode $i$ due to a flow at location $\br$, which is propagated to the observation points at the surface. Each mode is identified by its harmonic degree $l$ and radial order $n$. See Appendix~\ref{appendixoptimization} for a recap of the definitions of $J_{ij}$ and $\bZ^{ij}$.

\section{FAST COMPUTATION OF SPHERICAL BORN KERNELS}
\label{secfast}

As the numerical evaluation of Equation \eqref{eqcalckernel} is quite costly for obtaining an accurate kernel (about 2 days on 32 CPU cores for a three-dimensional kernel covering only about 5~\% of the solar volume with a coarse spatial grid), %
it is necessary to optimize its computation. This requires a little more mathematical detail, which the reader may well skip in order to understand the main results of this paper.

\subsection{3D Kernels}
\label{secopt3d}

We here propose an approach in which we 
make use of the separability of the eigenfunctions into a horizontal and a radial dependence. This is a consequence of the separation of variables performed when solving the oscillation equations (e.g., \citealp{Aerts2010}). The horizontal dependence of the eigenfunction is then also separable from the radial order of the mode. 
This allows us to rewrite the kernel formula \eqref{eqcalckernel} (see Appendix \ref{appendixoptimization} for details)
\begin{align}
    K_m(\br_1,\br_2;\br) 
				&=  \rho_0(r)  \, \sum_{\bar l, l} \, \sum_{k=r,\theta,\phi} \, q_{m,k,\bar l,l}(\Omega_1,\Omega_2,\Omega) \nonumber \\
				& \quad \times \, T_{m,k,\bar l,l}(r;\br_1,\br_2)     + \Big( {\bf 1 \leftrightarrow 2 }\Big)^* 
				. \label{eqcalckernelsuperfast}
\end{align}
Here, the quantity $q$ 
includes the horizontal dependence of the kernel from $\bZ^{ij}$. The quantity $T_{m,k,\bar l,l}$ includes the sum over $n$ and $\bar n$, the factor $J_{ij}$ as well as the $n,\bar n$-dependent factors and the radial dependence from $\bZ^{ij}$ in Equation~\eqref{eqcalckernel}. If $T_{m,k,\bar l,l}$ is evaluated before the main loop over $l$, $\bar l$, and the spatial grid in Equation~\eqref{eqcalckernelsuperfast}, the computational burden of the sum over $n$ and $\bar n$ does not play a considerable role in the total computation time anymore.

In practice, for a given $l$, the number of radial orders to be summed over is about 10 for probing the deep solar interior. The approach presented here thus decreases the computation time by about two orders of magnitude.

Computing a full 3D sensitivity kernel using harmonic degrees $22\leq l  \le 170$ and a spatial grid with $75\times1800\times3600$ grid points ($r\times \theta \times \phi$, covering depths until $0.6R_\Sun$ and the complete horizontal domain), therefore, takes about 1 day on 8 CPUs. 
A complete set consists of one kernel per travel distance, which can be reprojected to different latitudes and used to obtain, e.g., kernels for point-to-arc travel times, see examples presented in Figure~\ref{figkernels}. Computing such a set with 126 kernels as used in Sections~\ref{secfwdcomp} - \ref{secinversions} takes about 3 weeks on 96 CPUs.
%
%
%

Very recently, \citet{Gizon2016} developed an alternative approach for computing Born kernels. This approach is based on numerically solving for the Green's functions for individual modes and individual frequencies. It allows for a rather flexible computation of kernels for different quantities such as sound-speed, density, flows, and damping properties, as well as a possible inclusion of axisymmetric perturbations to the model.

While a detailed comparison of both methods is left to further studies, we consider as an example the computation of the radial solar model case from Table~2 in \cite{Gizon2016} using the frequency resolution from this study (about 92,000 grid points spread over a range of $3\,\rm{mHz}$, see also \citealp{Boening2016}) and 149 modes. This would thus take about eight times longer than using our method. In general, the computation time for our method scales with the number of modes squared, and the method of \cite{Gizon2016} scales with the number of modes times the number of frequency bins used. As a very fine frequency resolution is necessary to adequately sample the frequency domain in the case of deeply-penetrating modes \citep{Boening2016}, our method is thus advantageous for computing kernels for filtered observations of deep flows, and theirs for computing kernels using many modes or a coarser frequency resolution.

\subsection{2D Integrated Kernels}

Furthermore, it is possible to reach an even greater optimization when computing individual kernels which were integrated over the azimuthal coordinate $\phi$. For inferring flow fields which are rotationally symmetric such as meridional flow or differential rotation, one can rewrite Equation~\eqref{eqkernelgoal} as
\begin{align}
\EE[\delta \tau] 
	&= \iint \bv(r,\theta) \cdot \vec{\mathbb{K}}(\br_1,\br_2;r,\theta) \, r \, \id \theta \, \id r , \label{eqintkernelgoal}
\end{align}
where the integrated Kernel $\vec{\mathbb{K}}$ is defined as
\begin{align}
    \vec{\mathbb{K}}(\br_1,\br_2;r,\theta) &=  r\sin\theta \, \int   {\bK}(\br_1,\br_2;r,\theta,\phi) \,  \id \phi. 
\end{align}
We thus have
\begin{align}
    \mathbb{\bm K}_m(r,\theta) 
				&= r \rho_0(r) \sin\theta \, \sum_{k=r,\theta,\phi} \, \sum_{\bar l, l} \, Q_{m,k,\bar l,l}(\Omega_1,\Omega_2,\theta) \nonumber \\
				& \quad \quad \quad \times \, T_{m,k,\bar l,l}(r;\br_1,\br_2)     + \Big( {\bf 1 \leftrightarrow 2 }\Big)^* \label{eqcalcintkernelsuperfast} ,
\end{align}
where $Q_{m,k,\bar l,l}(\Omega_1,\Omega_2,\theta)$ is the integral over the $\phi$-coordinate of the variable $q$ in Equation~\eqref{eqcalckernelsuperfast}.

As the numerical evaluation of the integral of $q$ over $\phi$ is independent of the radial coordinate, for each latitude, the integration over $\phi$ and the loop over the radius become independent. As a consequence, the computation of an integrated kernel scales with the maximum of the number of radial and azimuthal grid points rather than with their product. For the example presented in Section~\ref{secopt3d}, the computation of an integrated 2D kernel is thus about 75 times faster compared to its 3D version. 
%
A drawback of this approach, however, is that 2D integrated kernels cannot be projected to different latitudes due to the nature of the coordinate system used. Therefore, this approach turns out to be advantageous if a smaller number of kernels is to be computed.

\subsection{Point-to-Arc and other Geometries}

It is possible to extend the previous approach for computing 2D integrated kernels to the case of travel times which were averaged over multiple points, e.g. in a point-to-arc geometry, if all individual point-to-point measurements were obtained for the same travel distance $\Delta_{1,2}$. Indicating with $\bar{\mathbb{\bm K}}$ and $\bar Q$ averages of $\bK$ and $Q$ over all pairs of observation points, we have
\begin{align}
\bar{\mathbb{\bm K}}(r,\theta) 
				&= r \rho_0(r) \sin\theta \, \sum_{k=r,\theta,\phi} \, \sum_{\bar l, l} \, \bar Q_{m,k,\bar l,l}(\Omega_1,\Omega_2,\theta) \nonumber \\
				& \quad \quad \times \, T_{m,k,\bar l,l}(r; \Delta_{1,2})     + \Big( {\bf 1 \leftrightarrow 2 }\Big)^* \label{eqcalckernelsuperfastsumobs}.
\end{align}

\section{BORN APPROXIMATION FORWARD MODEL COMPARED TO SIMULATED DATA}
\label{secfwdcomp}

In the following, we use a numerical simulation of helioseismic wave propagation in the solar interior by \citet{Hartlep2013}, which includes a standard single-cell meridional flow profile and which has been used before for validation purposes (see \citealp{Hartlep2013}, \citealp{Jackiewicz2015}, \citealp{Roth2016}, and references therein). As explained in \citet{Hartlep2013}, the flow was amplified by a factor of about 36 to a maximum amplitude of $500\,\rm{m\,s}^{-1}$ in order to increase the signal-to-noise ratio. As the simulation is linear, this increase does not alter the physics.

The radial displacement of the oscillations in the simulation at $r=R_\sun + 300\,\mathrm{km}$ are used as Dopplergrams. They are our input artificial data. Power spectra, cross-covariance functions, and travel times were obtained from this artificial data by \citet{Jackiewicz2015} using a time-distance measurement technique with phase-speed filters \citep{Kholikov2014}. 
In the following, these observational quantities are compared to the zero- and first-order models to be validated in this paper.

\subsection{Power Spectra}
\label{secpower}

\begin{figure*}%
\begin{center}

\includegraphics[width=0.33\textwidth]{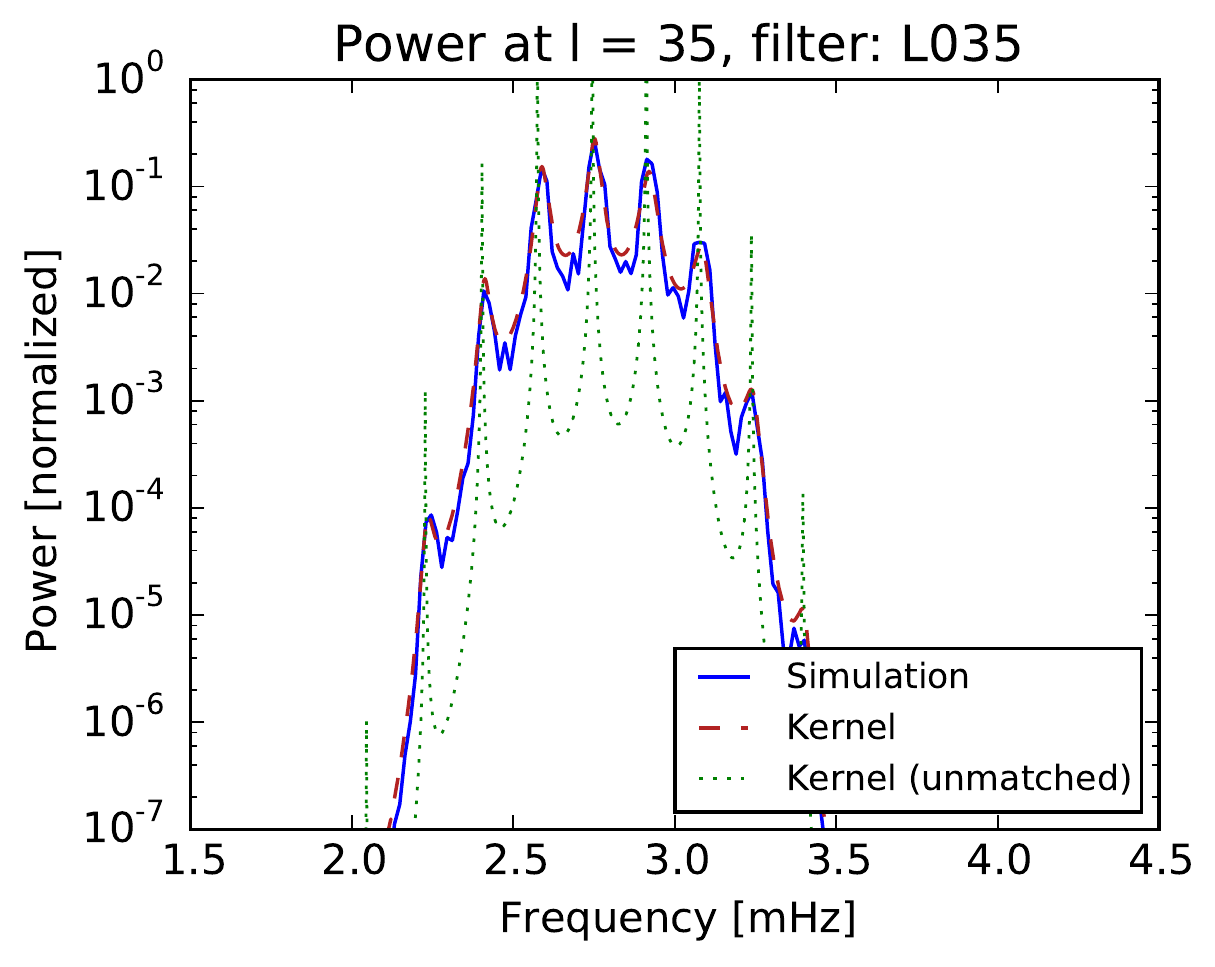}%
\includegraphics[width=0.33\textwidth]{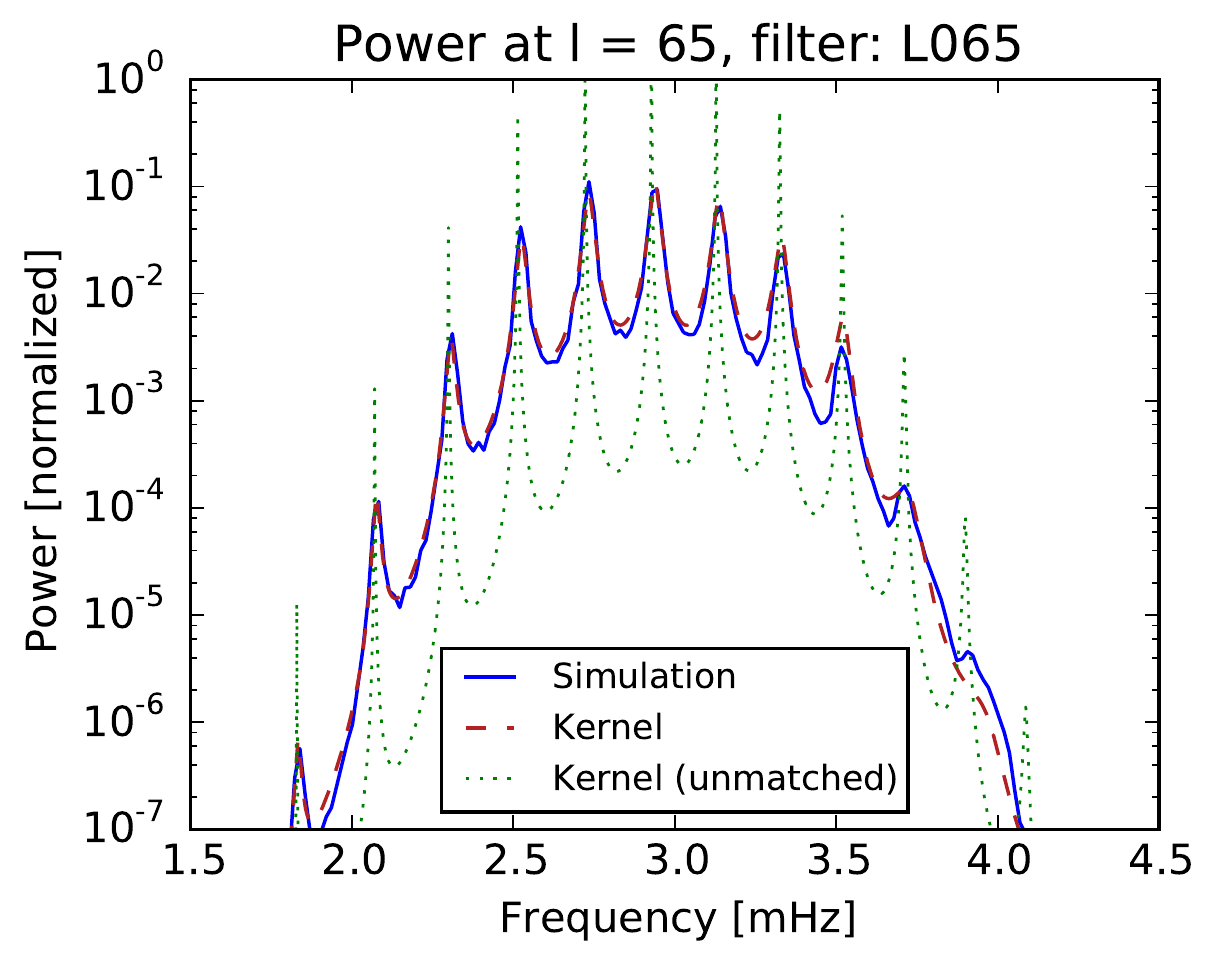}%
\includegraphics[width=0.33\textwidth]{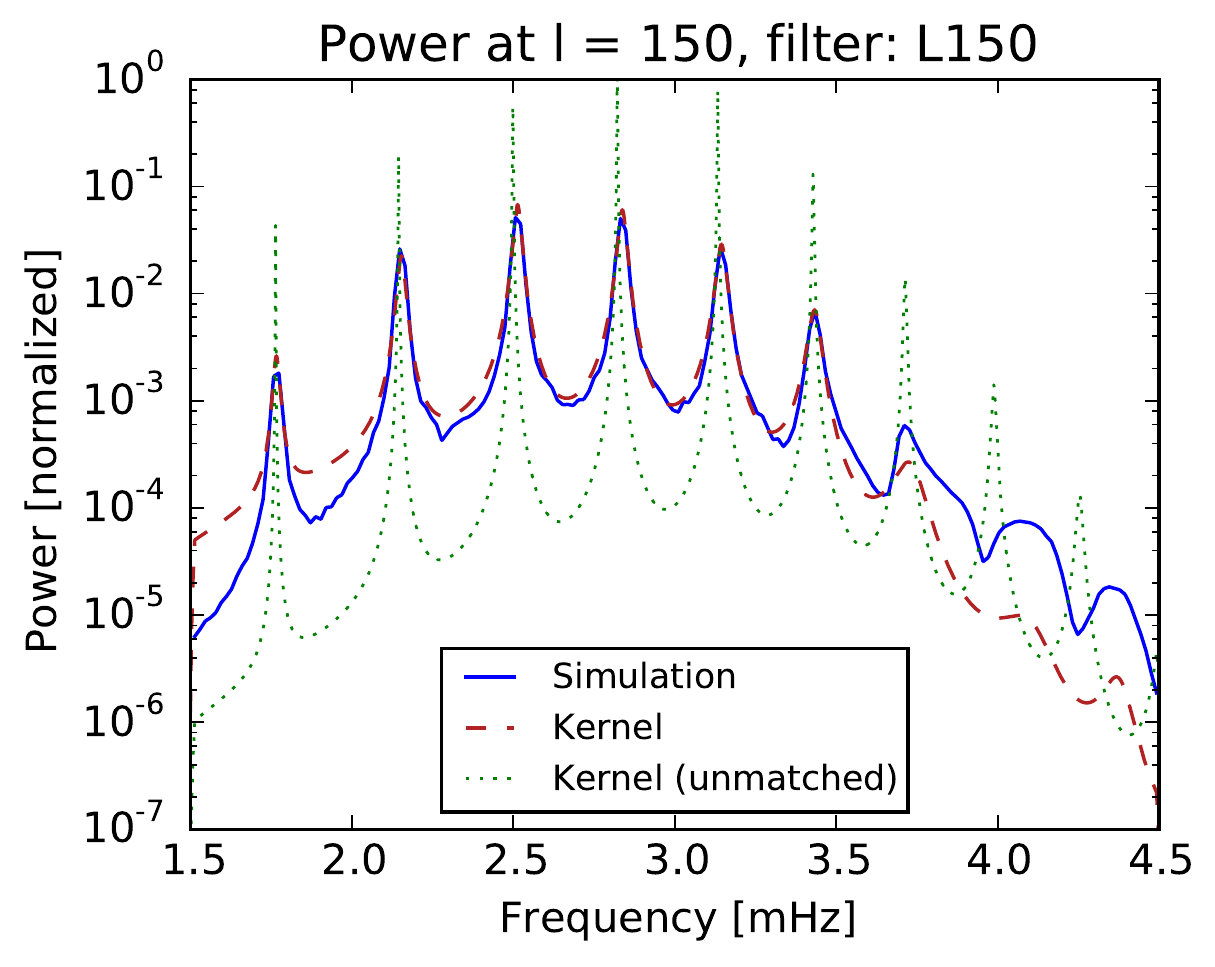}%

\includegraphics[width=0.33\textwidth]{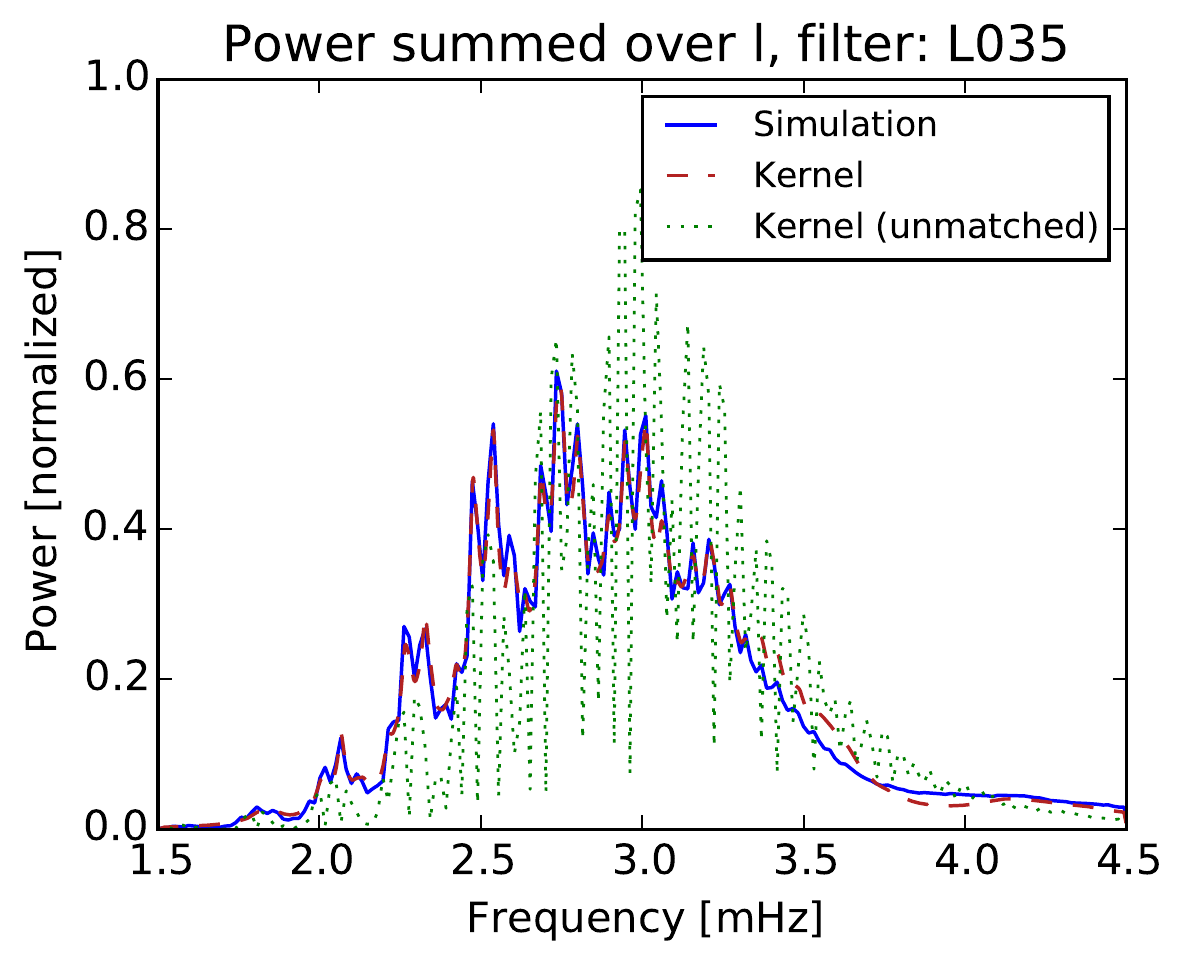}%
\includegraphics[width=0.33\textwidth]{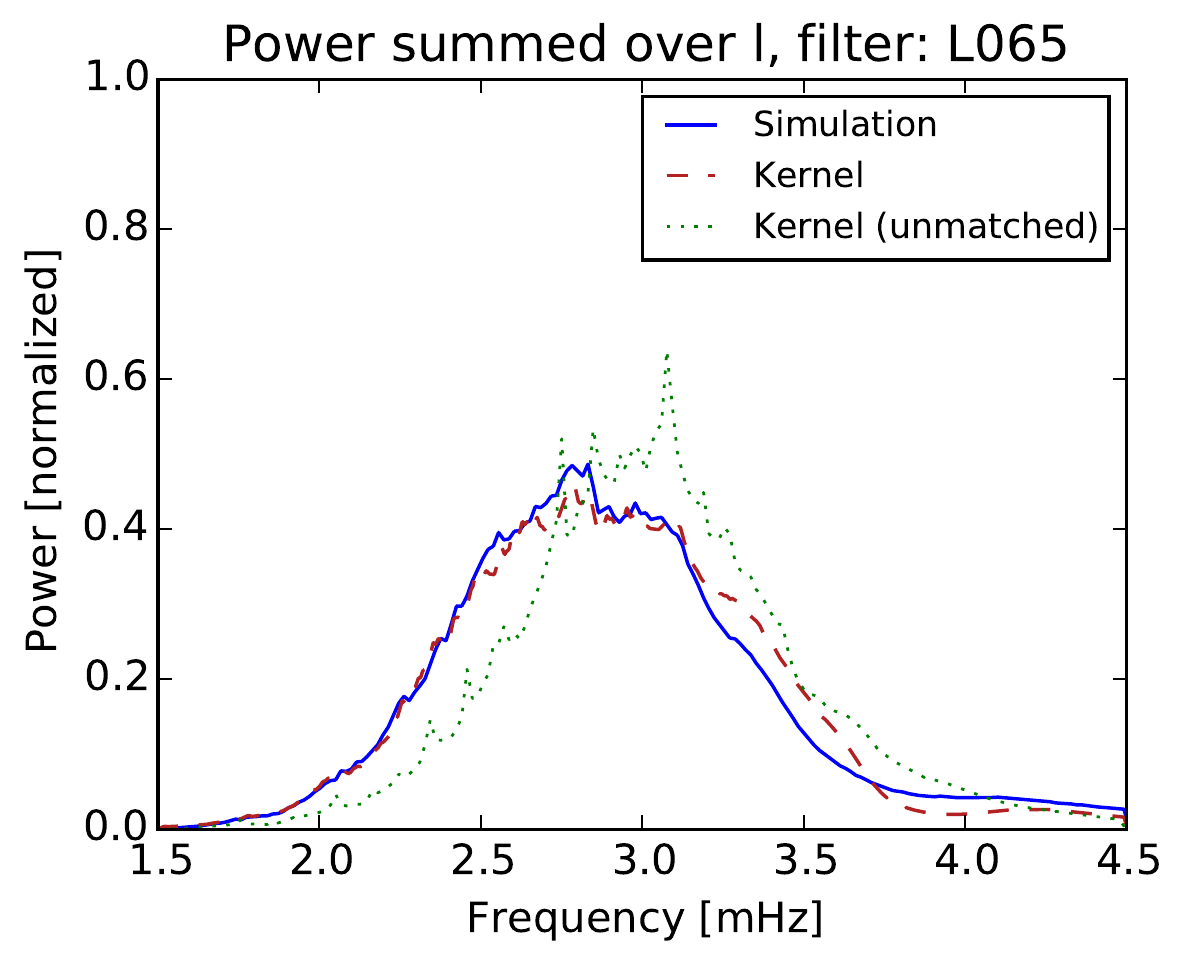}%
\includegraphics[width=0.33\textwidth]{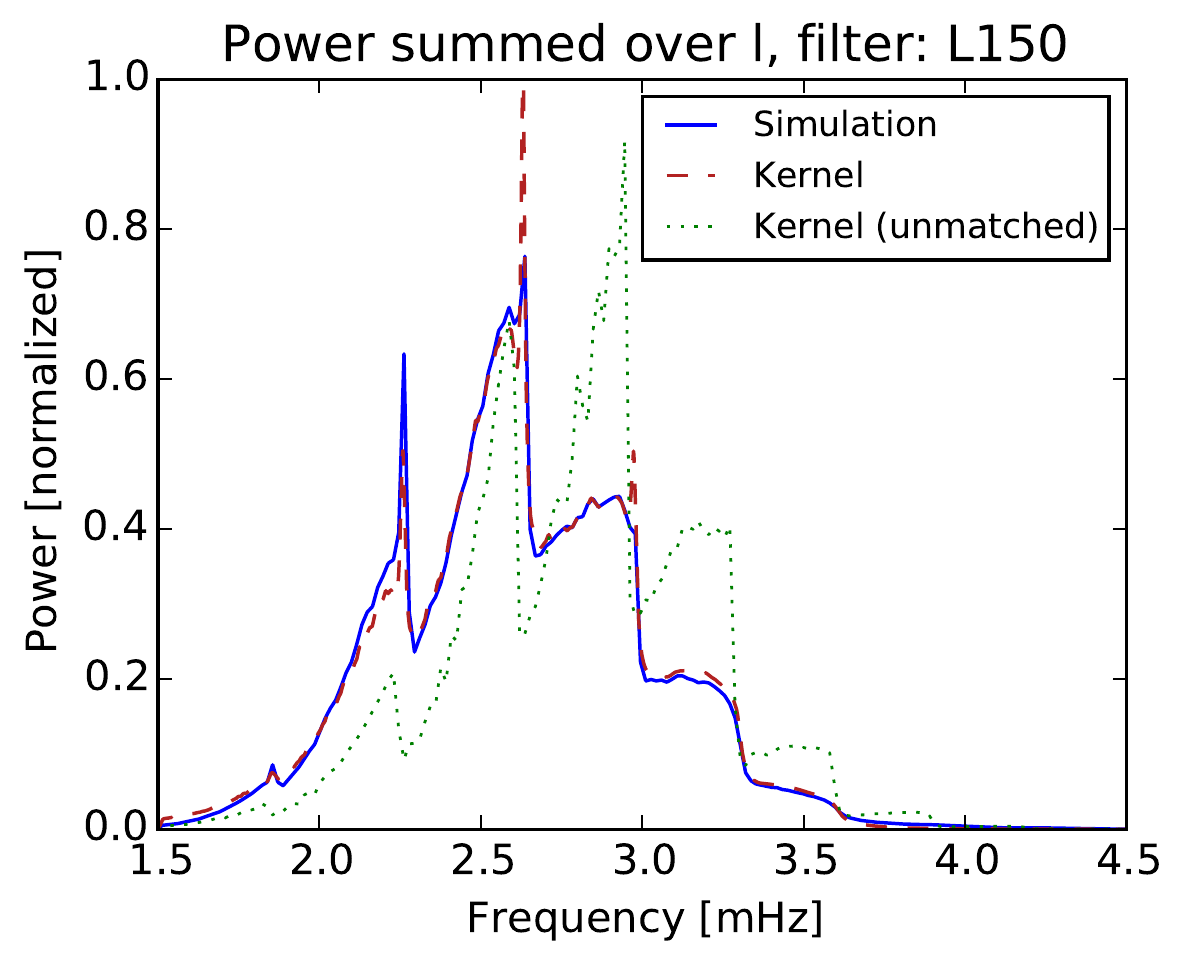}%

\end{center}

\caption{Comparison of filtered zero-order model power spectra with simulated data (blue lines) for three filters (different columns; for details on filters, see Table~\ref{tabfilters}). The model power spectra were obtained with two different sets of mode frequencies and damping rates (red dashed: mode frequencies and damping rates fitted to the simulated power; green dotted: using Model S eigenfrequencies and MDI damping rates). Top row: Cuts through the power spectra at the central harmonic degree of each filter. Bottom row: Power spectra summed over $l$. All power spectra were normalized to total power equal to one.\label{figpowercomp}}%
\end{figure*}

\begin{figure*}%
\begin{center}

\includegraphics[width=0.33\textwidth]{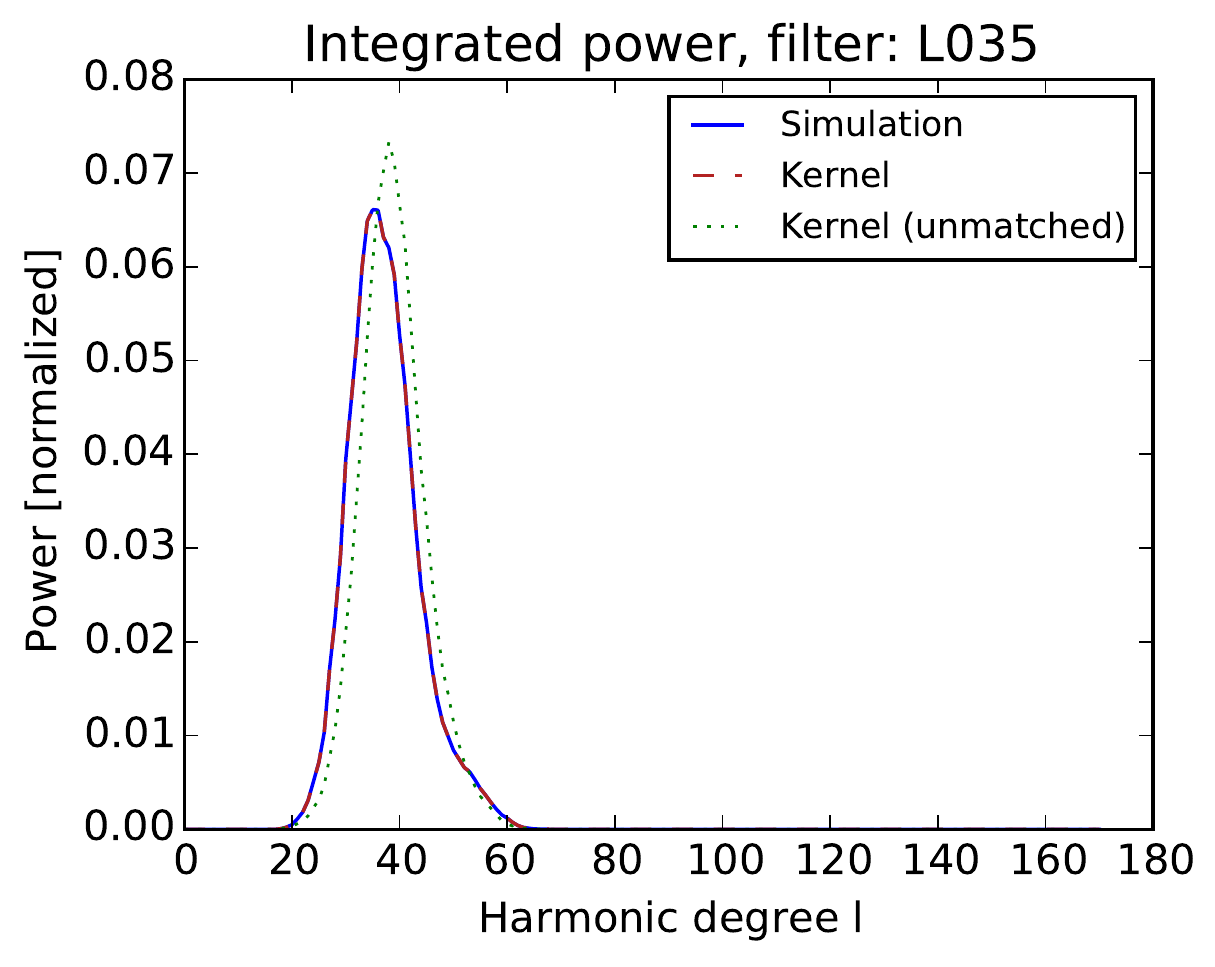}%
\includegraphics[width=0.33\textwidth]{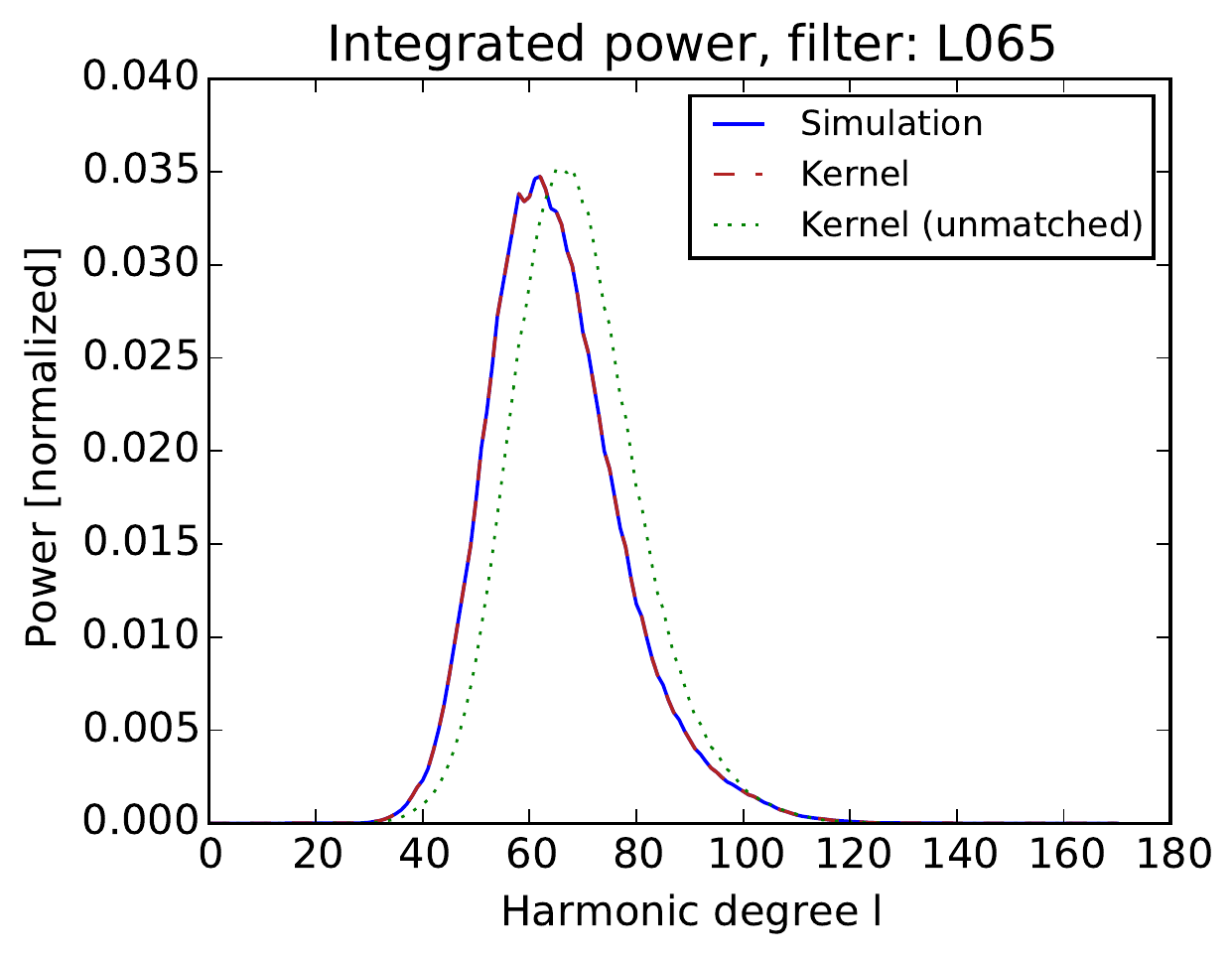}%
\includegraphics[width=0.33\textwidth]{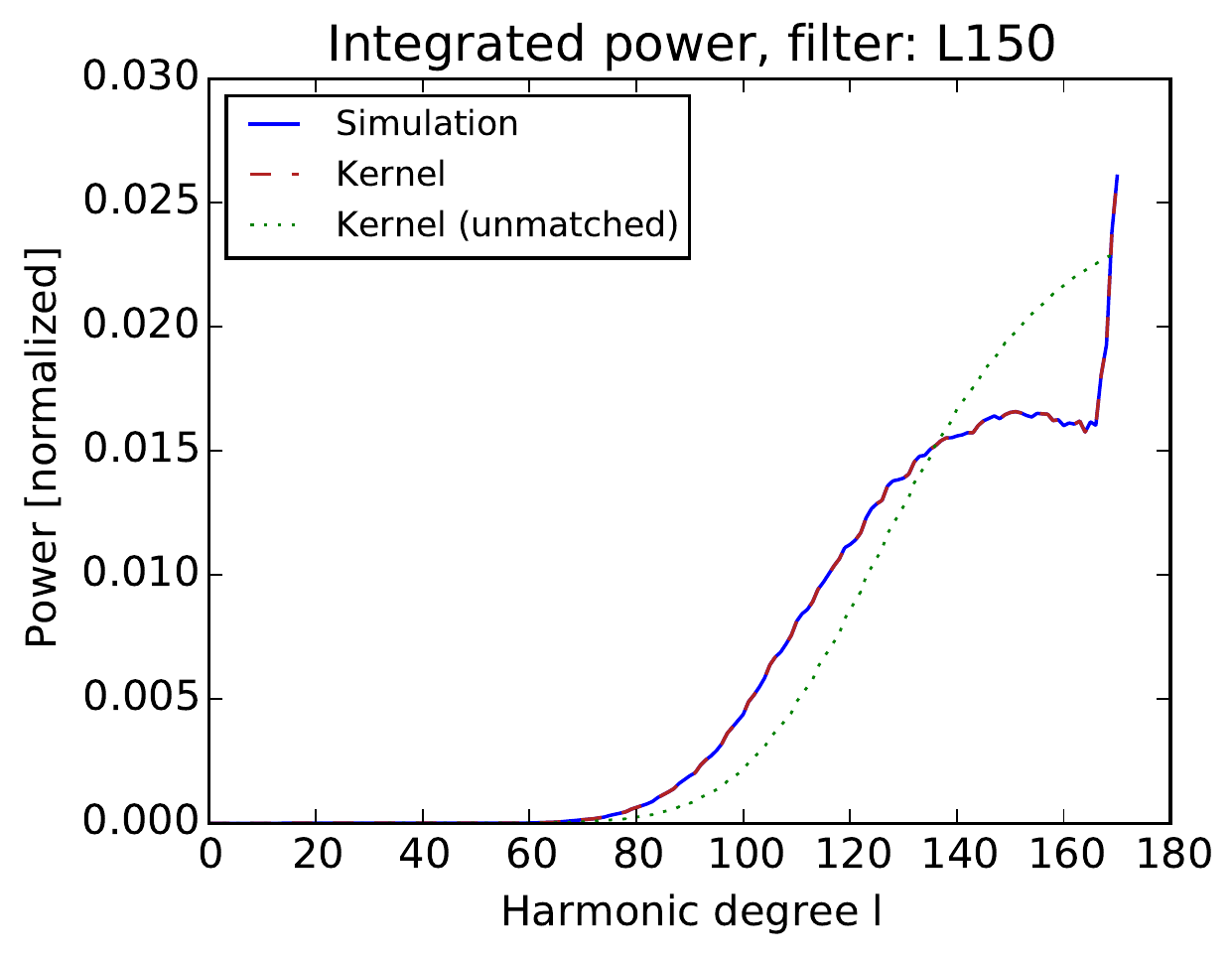}%

\end{center}

\caption{Comparison of the same power spectra as in Figure~\ref{figpowercomp}, integrated over frequency. The observational power spectra from simulated data (blue solid line) and model power spectra (red dashed, matched) are nearly identical. Unmatched power spectra are also shown (green dots).\label{figpowercomp2}}%
\end{figure*}

\begin{deluxetable}{cccccc}
\tabletypesize{\scriptsize}
\tablecolumns{6}
\tablecaption{Key Characteristics of Filtered Power Spectra.\label{tabfilters}}
\tablewidth{0pt}
\tablehead{
\multicolumn{2}{c}{} & \multicolumn{2}{c}{Matched P.} & \multicolumn{2}{c}{Unmatched P.} \\
\colhead{Filter} & \colhead{$v_{\rm{ph}}$} & \colhead{Mean $\nu$} & \colhead{Mean $l$} & \colhead{Mean $\nu$} & \colhead{Mean $l$} \\
\colhead{} & \colhead{$(\rm{km}\,\rm{s}^{-1})$} & \colhead{$(\rm{mHz})$} & \colhead{} & \colhead{$(\rm{mHz})$} & \colhead{}
}
\startdata
  L035  &  342.9 &   2.93  &   37.2 &   3.05  &   38.9 \\
  L040  &  308.5 &   2.93  &   41.5 &   3.05  &   43.4 \\
  L045  &  274.6 &   2.93  &   46.7 &   3.05  &   49.0 \\
  L050  &  248.8 &   2.92  &   51.8 &   3.05  &   54.5 \\
  L065  &  199.4 &   2.91  &   65.1 &   3.05  &   69.0 \\
  L080  &  158.7 &   2.90  &   82.8 &   3.05  &   88.5 \\
  L100  &  130.7 &   2.87  &  101.0 &   3.04  &  108.9 \\
  L120  &  106.9 &   2.78  &  121.5 &   2.96  &  130.1 \\
  L150  &   78.5 &   2.63  &  138.8 &   2.81  &  143.9 \\
  L170  &   30.5 &   2.16  &  157.3 &   2.24  &  156.4 \\
\enddata
\tablecomments{For each filter (left column, the last three digits represent the central harmonic degree of the filter), we summarize the central phase speed used (second column), as well as the power-weighted mean frequency and harmonic degree of the filtered zero-order power spectra which were matched to the observational one (third and fourth column) and not matched (fifth and sixth column). The values for the observational power spectra extracted from the simulation are identical to the matched case. Travel distances used for each filter can be identified in Figure~\ref{figfwdtts2d}.}
\end{deluxetable}

Filtered power spectra obtained from the simulation and from the zero-order model are compared in Figures~\ref{figpowercomp} and \ref{figpowercomp2}. Figure~\ref{figpowercomp} shows cuts through the power spectra at the central harmonic degree of each filter (top) and power spectra summed over harmonic degree (bottom). Figure~\ref{figpowercomp2} shows power spectra integrated over frequency.  See Table~\ref{tabfilters} for details on the filters used.

In the following, the simulated data is compared to our model using two different sets of mode frequencies and damping rates. As a first-guess case,
the zero-order model power spectrum, displayed as a green dotted line, was computed with frequencies from Model S and damping rates from MDI (``unmatched'' in the legends appearing in this paper). Damping rates were provided by J. Schou (2006, private communication). 

It can be seen that the peaks in the model power spectrum are systematically different from the ones in the simulation. The widths of the peaks are systematically smaller in the model power spectrum and the central frequencies of the peaks differ, especially at frequencies above about $3.8\,\mathrm{mHz}$ (best observable in the top right panel). 
As was discussed in \citet[Section 3.3]{Boening2016}, the shape of a cut through the model power spectrum matches a Lorentzian function centered at the input mode frequency and with the damping rate as its half width at half maximum. The difference in the location and shape of the peaks between simulation and model thus shows that the mode frequencies and damping rates from the simulated data are different from Model S frequencies and observed damping rates.

As the computation of sensitivity functions is strongly dependent on accurately modelling the data power spectra (e.g., \citealp{GB2002,Birch2004,Jackiewicz2007a,DeGrave2014QS,DeGrave2014Sunspot}), with real solar observations, one would use mode frequencies and damping rates as close to solar values as possible. In order to achieve a good match between model and data power spectra for the validation presented in this paper, we therefore fitted frequencies and damping rates to the simulated power spectrum. These fitted values were used as parameters for the zero-order power spectrum shown as a red dashed curve in Figures~\ref{figpowercomp} and \ref{figpowercomp2} (``matched'' in the legends). It can be seen that both the locations and the widths of the peaks of the simulated and zero-order modelled power spectra now match well (Figure~\ref{figpowercomp}, top). The amplitude of each peak, however, is not a free parameter in the zero-order model and therefore cannot be adjusted to the simulation.

Additionally, the source correlation time, a free parameter modelling the sources in \citet{Boening2016}, was fine-tuned individually for each matched filtered zero-order model power spectrum in order to obtain a power-weighted mean frequency identical to the one from the simulation. 
This is necessary for guaranteeing that the mean sensitivity of the kernel is correctly adjusted (\citealp{Jackiewicz2007a}, \citealp{Boening2016}). In addition, the power spectra for the kernel were corrected by an $l$-dependent factor which accounts for a different behavior of the frequency-integrated power in the simulation and in the zero-order model, see Figure~\ref{figpowercomp2}.
See also Table~\ref{tabfilters} for a summary of properties for the power spectra obtained with matched and unmatched frequencies and damping rates, for each individual filter used in this work.

In the following, we will compare simulated data to the zero- and first-order models obtained using both sets of frequencies and damping rates.

\subsection{Cross-Covariances}

Born approximation sensitivity functions in time-distance helioseismology as in \citet{Boening2016}, \citet{BG2007}, and \citet{GB2002} are obtained for a travel-time fit which involves a reference cross-covariance function.

On the data analysis side, it is advantageous to use some kind of disc- and time-averaged cross-covariance function as a reference. This assures that travel-times measured at individual locations at a certain point in time correspond to local changes compared to the mean solar cross-covariance for a particular travel distance.

\begin{figure*}%
\begin{center}

\includegraphics[width=0.33\textwidth]{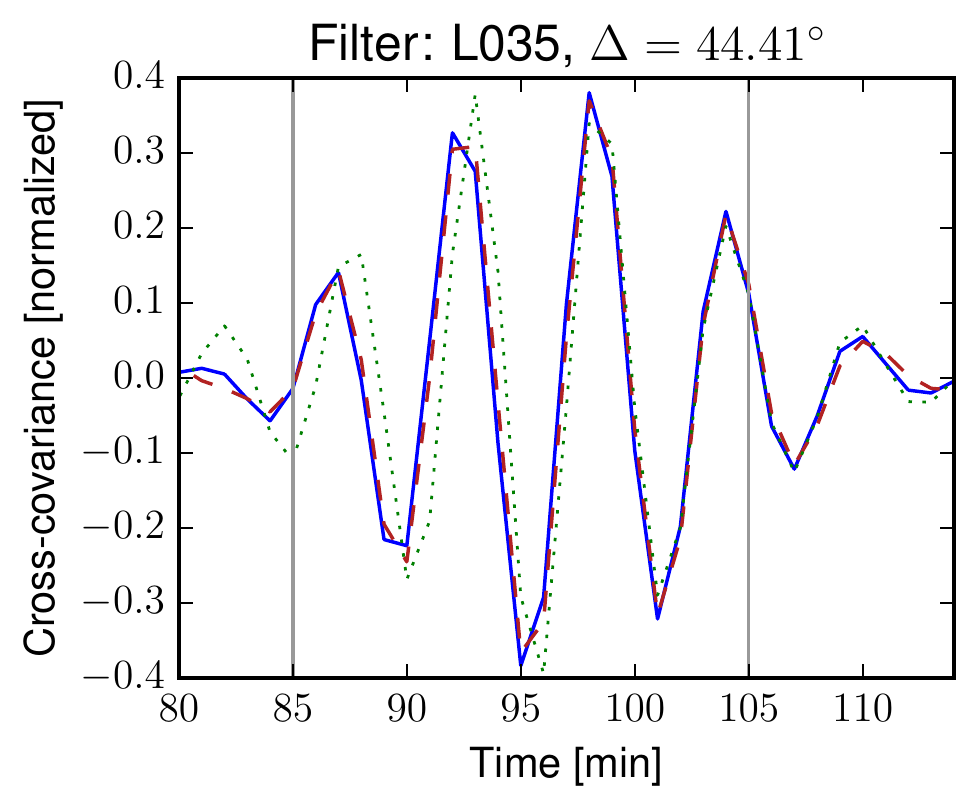}%
\includegraphics[width=0.33\textwidth]{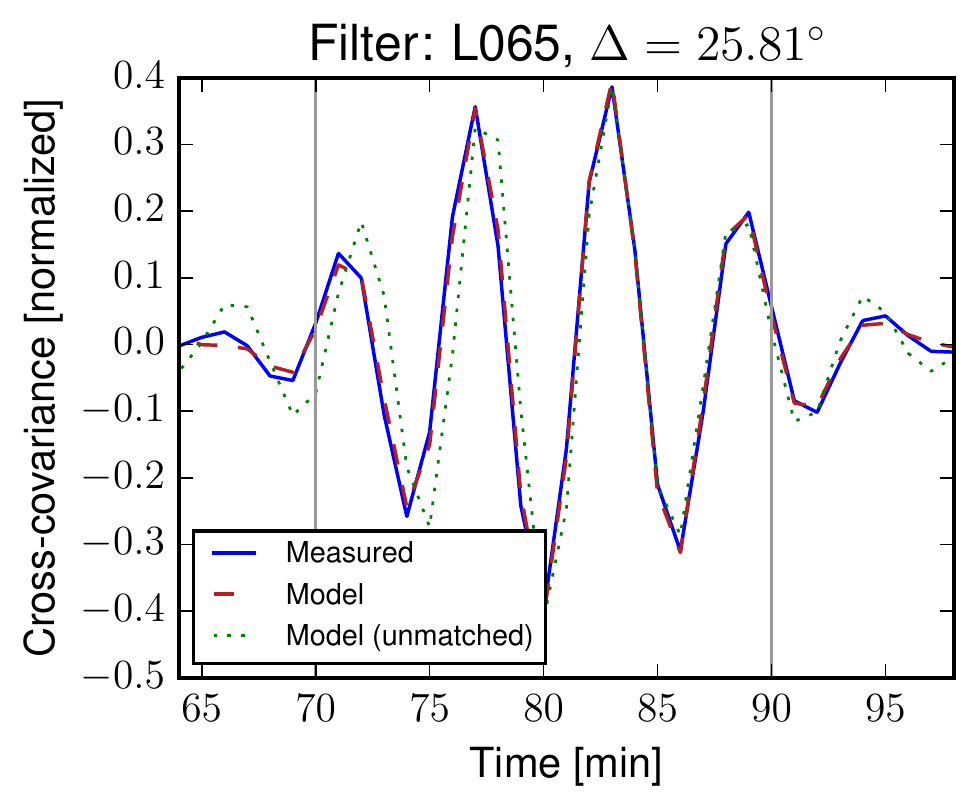}%
\includegraphics[width=0.33\textwidth]{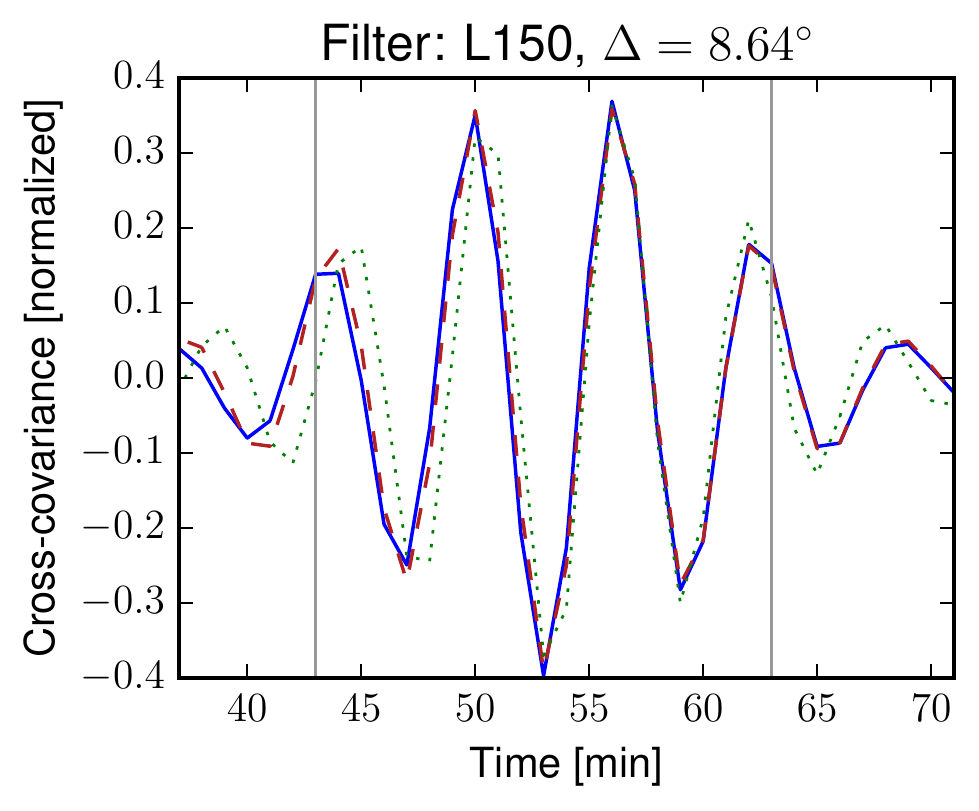}%

\end{center}

\caption{Reference cross-covariance functions for exemplary distances. We show disc-averaged mean cross-covariance functions from the simulation (blue) and the zero-order cross-covariance (red dashed: matched; green dotted: unmatched), which is used as reference cross-covariance function in the model. The perturbation to this reference in the presence of flows is modelled by the Born approximation. The vertical grey lines indicate the windows used for the GB04 travel-time fit. The whole displayed range was used for the Gabor fit.}%
\label{figcrefmatched}%

\end{figure*}

On the modelling side, the zero-order cross-covariance function is usually used as a reference function. Figure~\ref{figcrefmatched} shows a comparison of these reference cross-covariance functions for different exemplary travel distances. It can be seen that the model fits the data very well once the fitted frequencies and damping rates were used.

\subsection{Travel Times}
\label{sectts}

\begin{figure*}%

\includegraphics[width=0.33\textwidth]{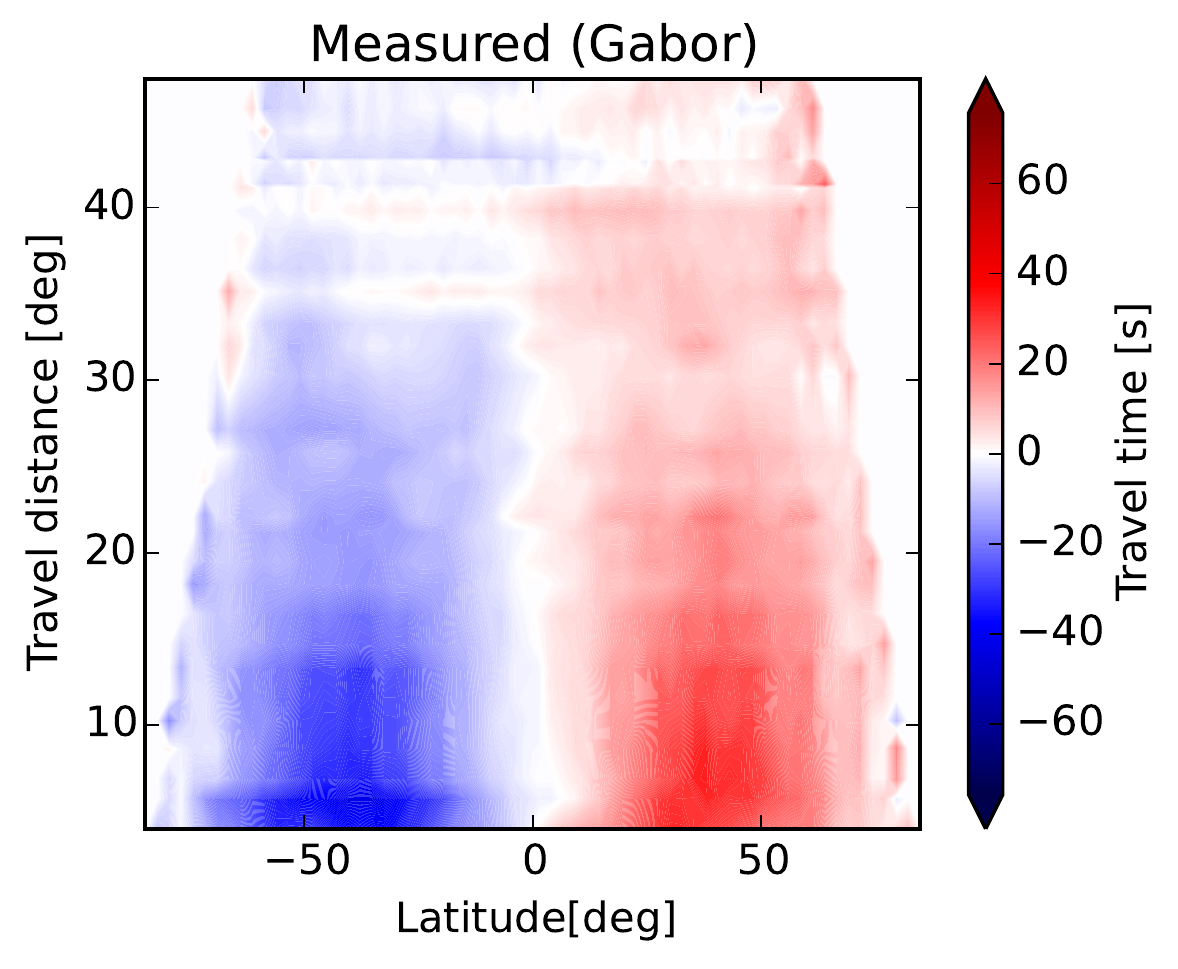}%
\includegraphics[width=0.33\textwidth]{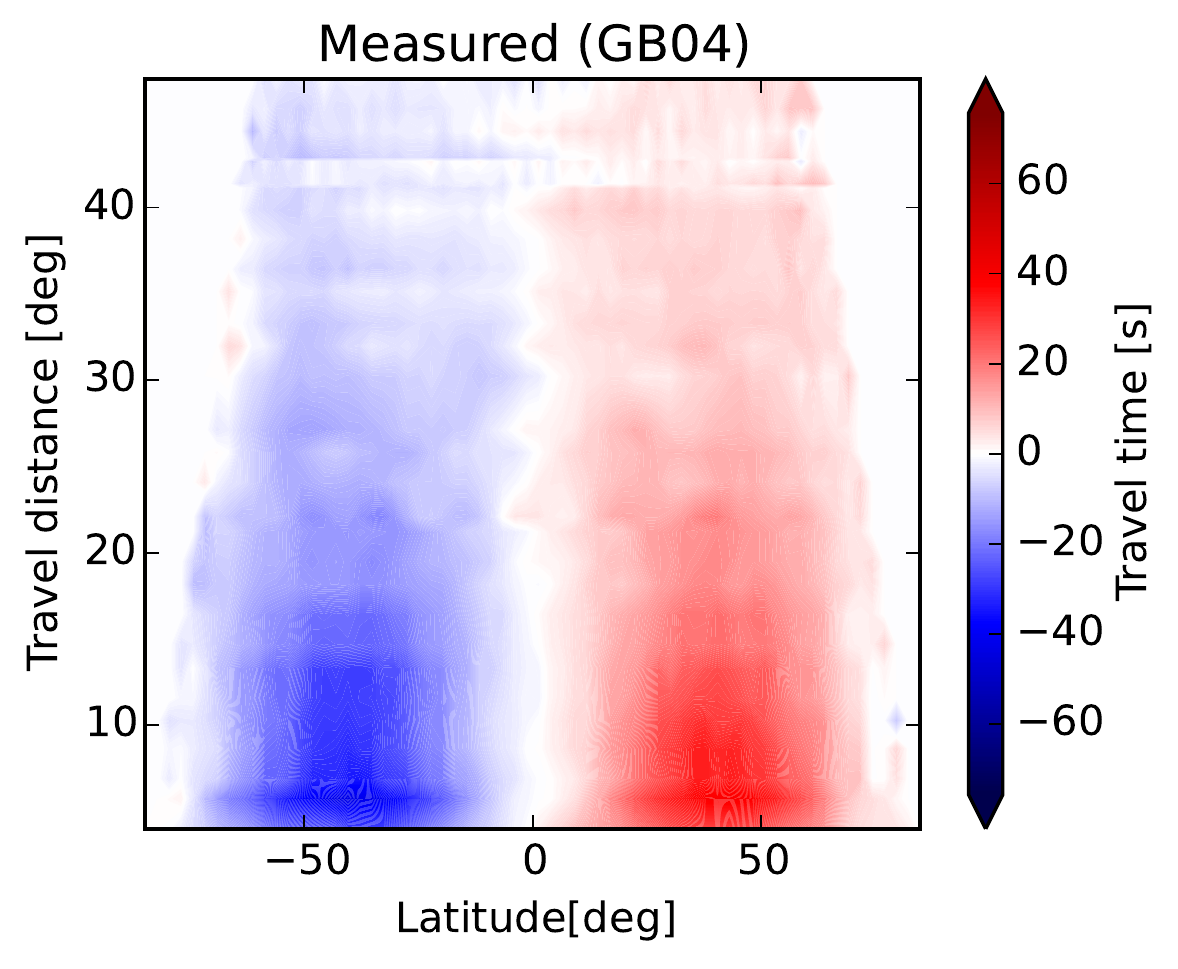}%
\includegraphics[width=0.33\textwidth]{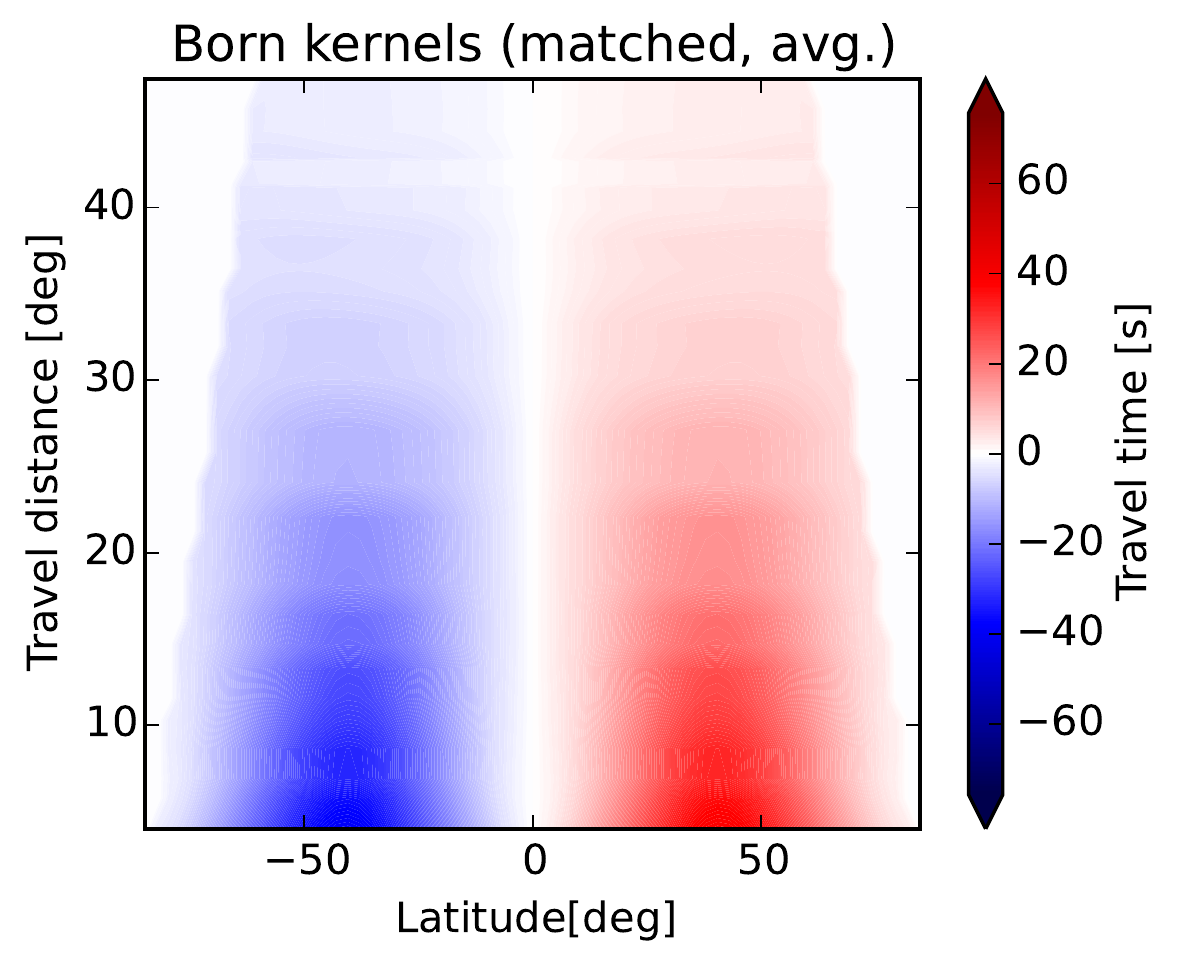}%

\includegraphics[width=0.33\textwidth]{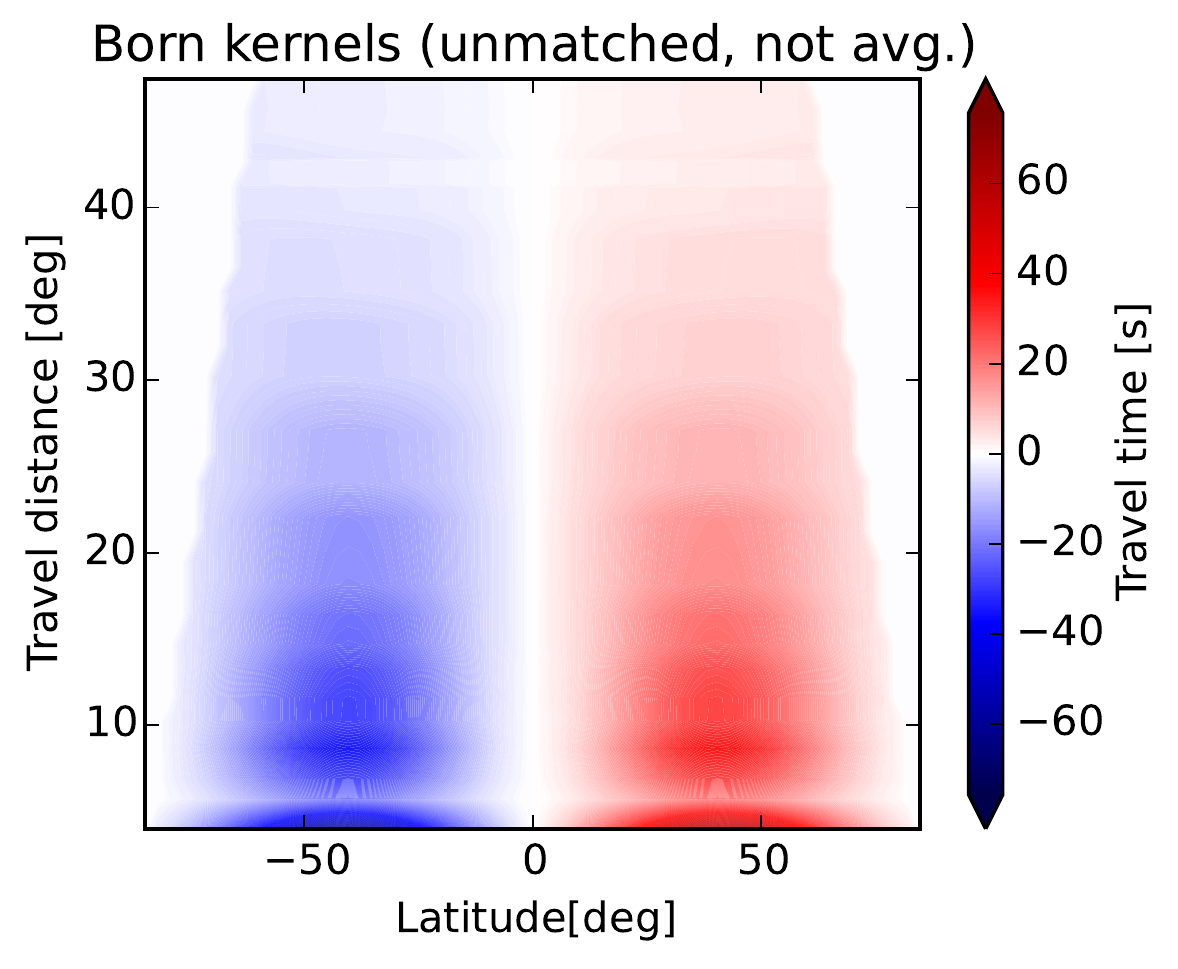}%
\includegraphics[width=0.33\textwidth]{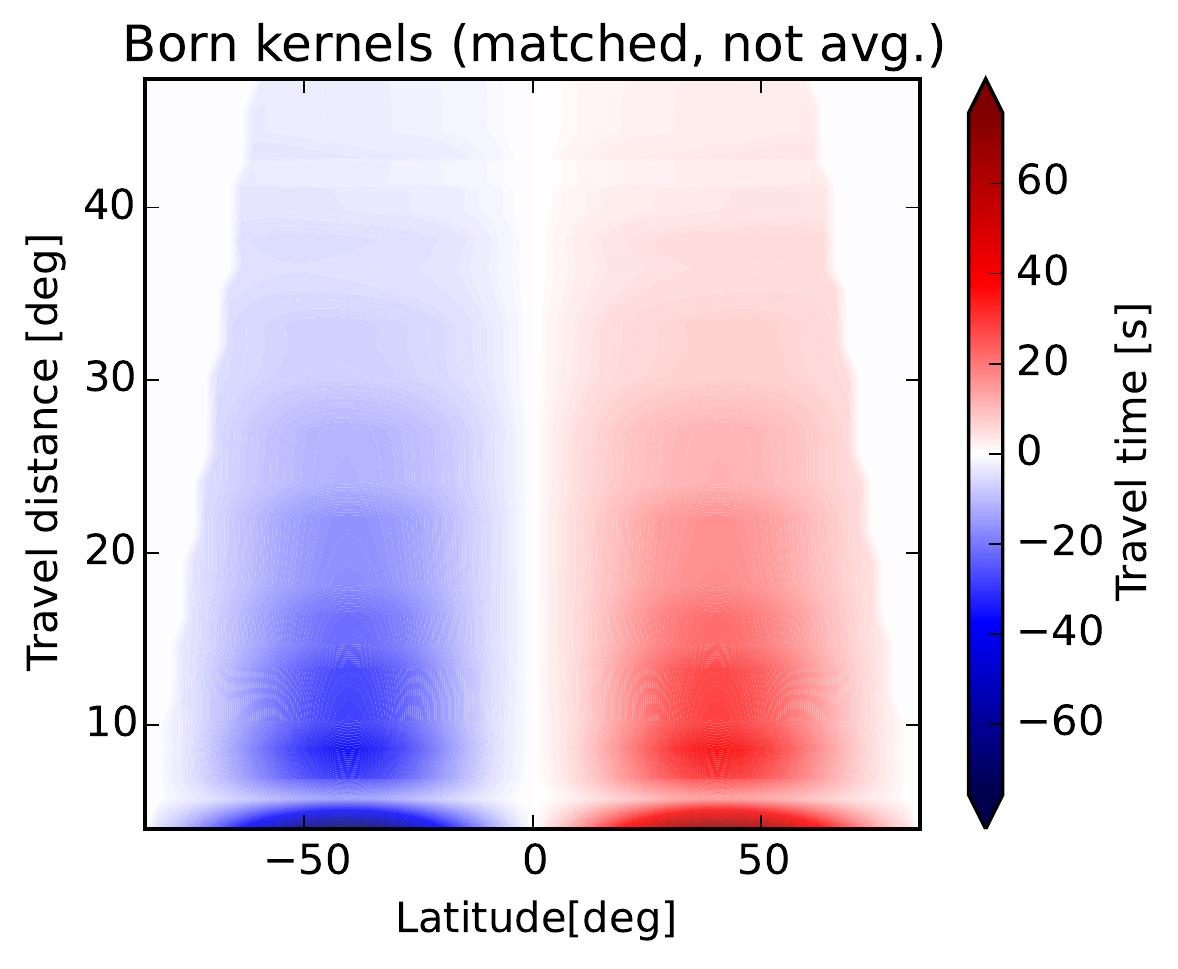}%
\includegraphics[width=0.33\textwidth]{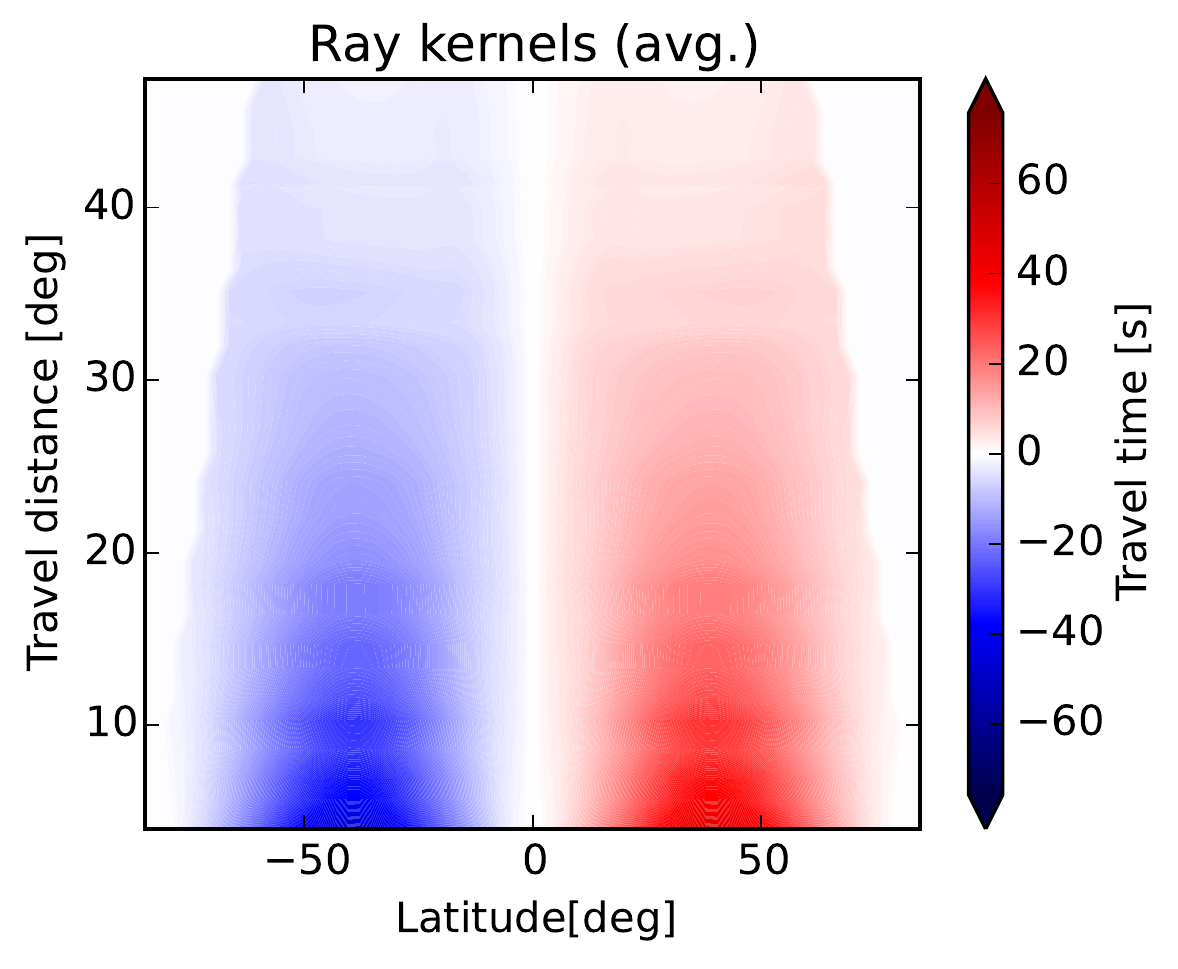}%

\caption{Travel times in point-to-arc geometry obtained using different methods. Top: The panels show measured travel times using a Gabor fit (left) and a GB04 fit (middle), as well as travel times from the Born kernels (right). Both measured and forward travel times were averaged over a set of distances and latitudes. Bottom: Forward travel times obtained using different sets of kernels, using the unmatched power spectrum and not averaging over distance and latitude (left, Born kernels), using the matched power spectrum and not averaging over distance and latitude (middle, Born kernels), and travel times obtained from ray kernels (right) averaged in the same way as the travel times displayed in the top row.
}%
\label{figfwdtts2d}%
\end{figure*}

Travel times fitted to the cross-covariances obtained from the simulated data can be compared to forward-modelled travel times, which can be obtained from the Born kernels using Equation~\eqref{eqkernelgoal} or \eqref{eqintkernelgoal} as the underlying flow field is known.

Figures~\ref{figfwdtts2d} and \ref{figfwdtts1d} show a comparison of such travel times. The displayed measured travel-times were obtained by rebinning an original set of travel times for 598 central latitudes and 126 distances to 66 latitudes and 30 distances. The corresponding averaging was applied to the kernels used for the forward travel times marked with ``avg.'' in Figures~\ref{figfwdtts2d} and \ref{figfwdtts1d}. For the forward-modeled travel times marked with ``not avg.'', however, the kernels have not been averaged. Instead, we have just computed one center-to-arc kernel for each of the 66 times 30 distances. Furthermore, we also show forward travel-times from kernels obtained with an unmatched power spectrum.

In Figure~\ref{figfwdtts2d}, it can be seen that the travel times from the kernels generally fit very well to the measured ones. The forward-modelled travel times, which were obtained using properly averaged Born kernels and a matched power spectrum (top right panel), also reproduce some of the jumps visible in the measured travel times as a function of distance at mid-latitudes. These jumps are introduced by a change of phase speed from filter to filter. It can also be seen that there do not seem to be significant differences between the Gabor and GB04 fits. 

Horizontal cuts through the panels displayed in Figure~\ref{figfwdtts2d} are shown in Figure~\ref{figfwdtts1d}, where the match between forward-modelled and measured travel times can be inspected in more detail. It can be seen that forward-modelled travel times fit the measured ones within the measurement errors for most distances. The agreement is particularly good for travel distances of about 8-20 degrees with measurement errors increasing with increasing travel distance.

At very low travel distances (about $\Delta<6\degr$), it can be seen that the forward-modelled travel times do not fit the measured ones. The reason for this effect lies in the fact that the simulated data only incorporates modes with harmonic degree $l\leq 170$. This cut-off acts as an additional filter at a harmonic degree at which the filter with the lowest phase speed (L170) is centered. The modes selected by the filter thus do not form a proper wave packet, which may lead to artifacts in the cross-correlations (see, e.g., \citealp{Zharkov2006}).

Additionally, it is noteworthy that the quality of the match of the power spectrum (compare matched vs. unmatched) does not have a large effect on the travel times at higher travel distances ($\Delta > 8\degr$). At smaller travel distances ($\Delta < 8\degr$), however, one can observe a substantial effect of the quality of the match of the power spectrum on the travel times. We note here that this conclusion may be different for other flow models. We also note that the absence of a proper averaging of multiple kernels in distance and latitude has a considerable effect for smaller travel distances ($\Delta < 10\degr$) but seems to vanish for the largest distances in the case of the flow model considered here.

For comparison, forward travel times using ray kernels are also shown in Figures~\ref{figfwdtts2d} and \ref{figfwdtts1d}. It can be seen that the magnitude of the ray kernel travel times is, in general, very similar to those obtained using Born kernels. However, Born kernels seem to better reproduce some features in the measured travel times, e.g., jumps in the magnitude of the travel times from one filter to another such as from $\Delta=22.1\degr$ to $\Delta=24.1\degr$ (see the middle panel in Figure~\ref{figfwdtts1d}). 
As the given flow model varies on rather large length scales, a relatively good performance of the ray kernels is expected. This may be different for meridional flow models which vary on smaller length scales, see Section~\ref{secintro}.

We also note here that the measured travel times obtained using the filter with the highest phase speed (L035) encounter a spurious constant offset, see Figure~\ref{figfwdttsnooffset}. This offset was corrected for by substracting the mean travel time at each distance. Such an offset is not present in real solar data (\citealp{Kholikov2014}, \citealp{Jackiewicz2015}). In \citet{Jackiewicz2015}, this offset was not noticed in the simulated data, as a few distances had been dropped from the analysis, which resulted in an equivalent correction. The origin of this offset is not completely understood but may be connected to the relatively coarse spatial resolution of the simulated data (256 pixels on 180 degrees).

\section{IMPACT OF THE RADIAL FLOW COMPONENT}
\label{secradial}

\begin{figure*}%

\includegraphics[width=\textwidth]{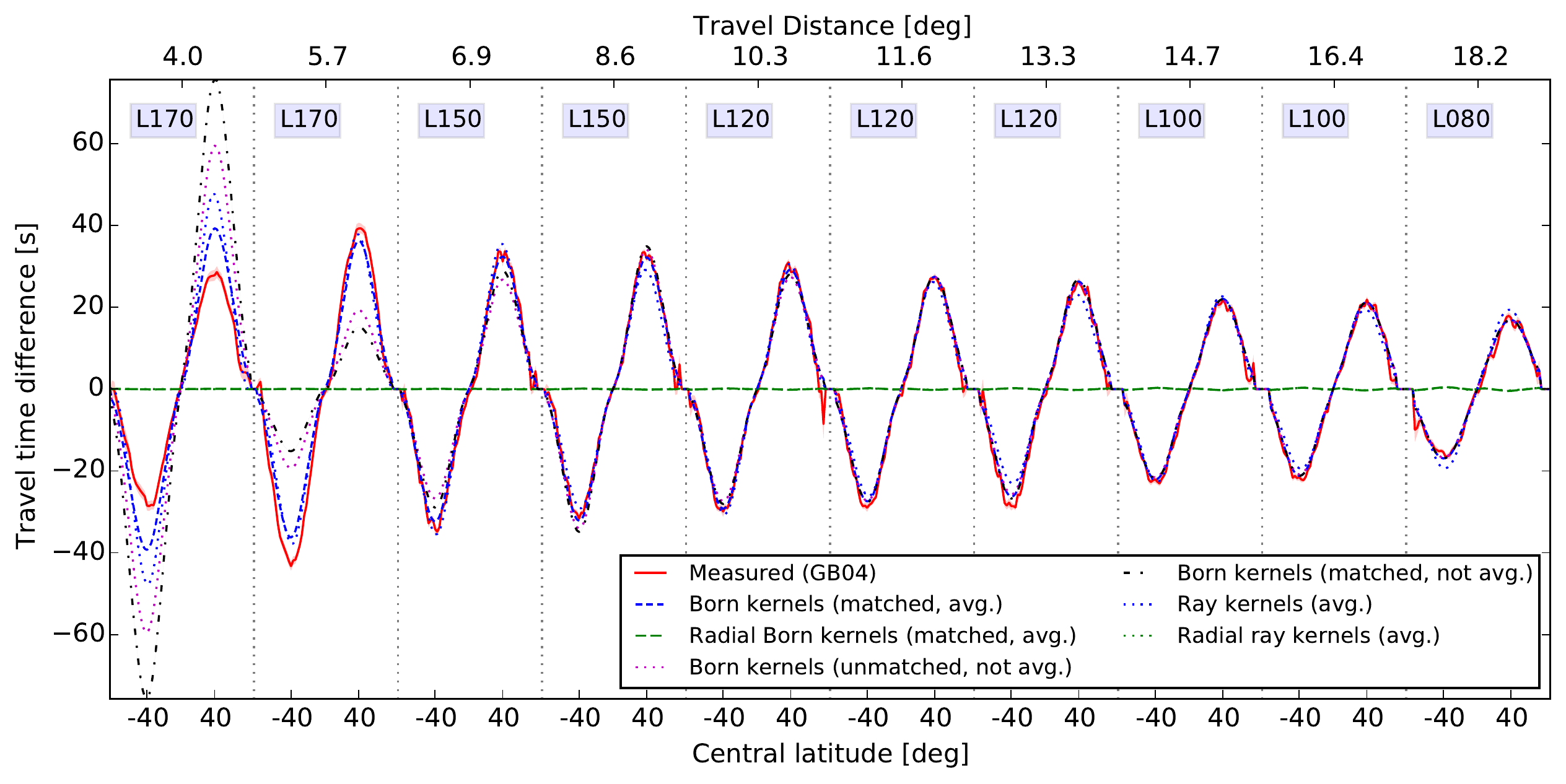}%

\includegraphics[width=\textwidth]{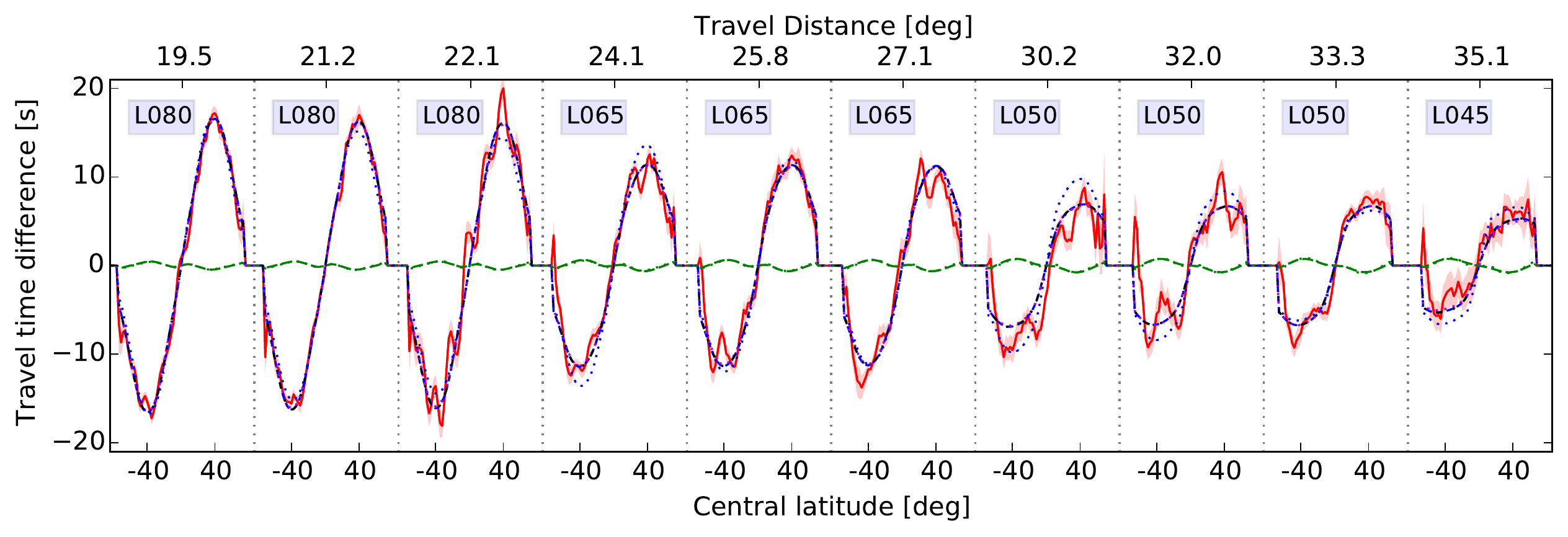}%

\includegraphics[width=\textwidth]{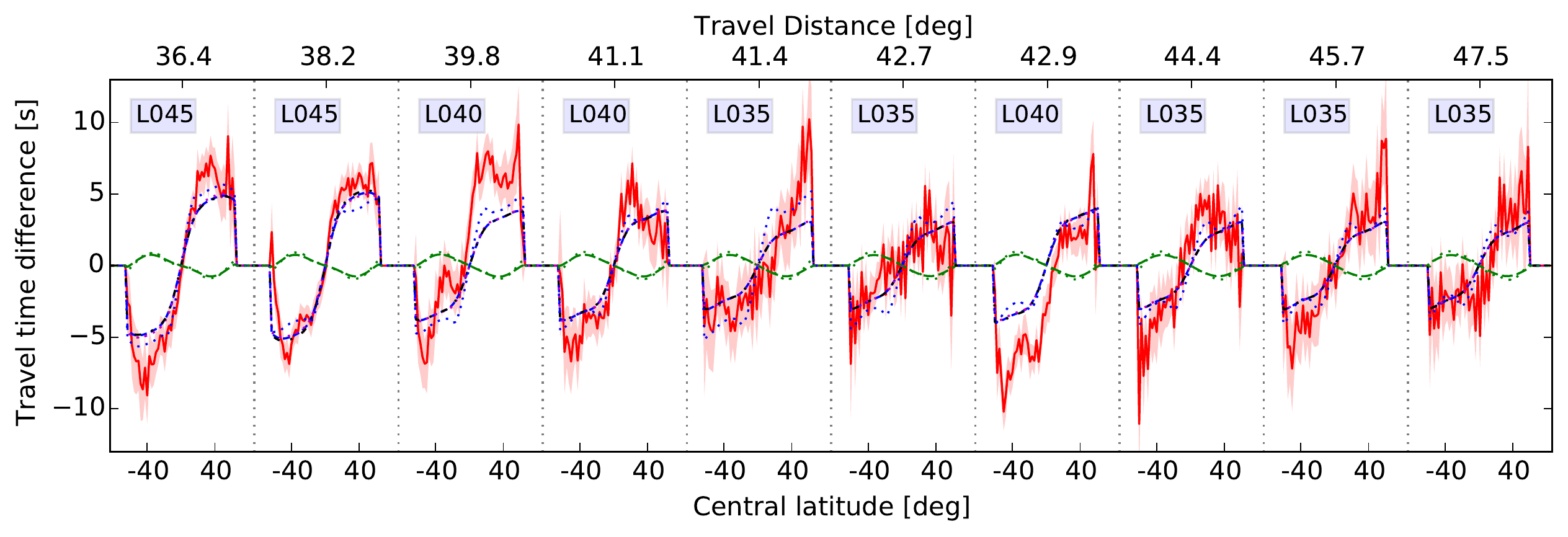}%

%

\caption{Comparison of forward-modelled (all dashed, dotted, and dashed-dotted lines) and measured travel times (solid red) including measurement errors (light red regions). All travel times shown in this figure, apart from the impact of the radial flow component on the travel times (green), are also shown in Figure~\ref{figfwdtts2d}. For all travel distances (upper x-axis), delimited by vertical grey lines, the travel times are plotted as a function of the central latitude of the observation points (bottom x-axis). At each distance, the filter applied is indicated in a blue box, see also Table~\ref{tabfilters}.}%
\label{figfwdtts1d}%
\end{figure*}

\begin{figure*}%

\includegraphics[height=1.6in]{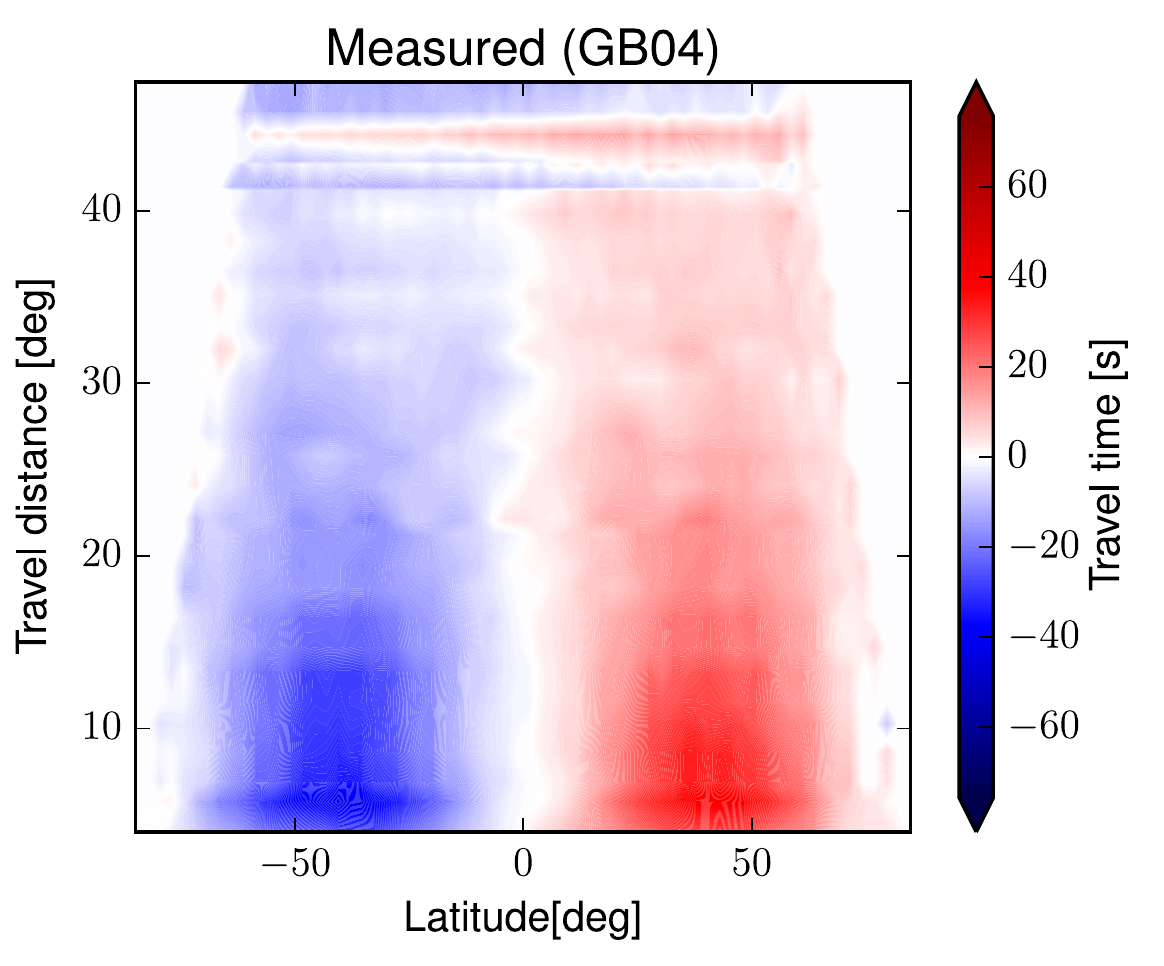}%
\includegraphics[height=1.6in]{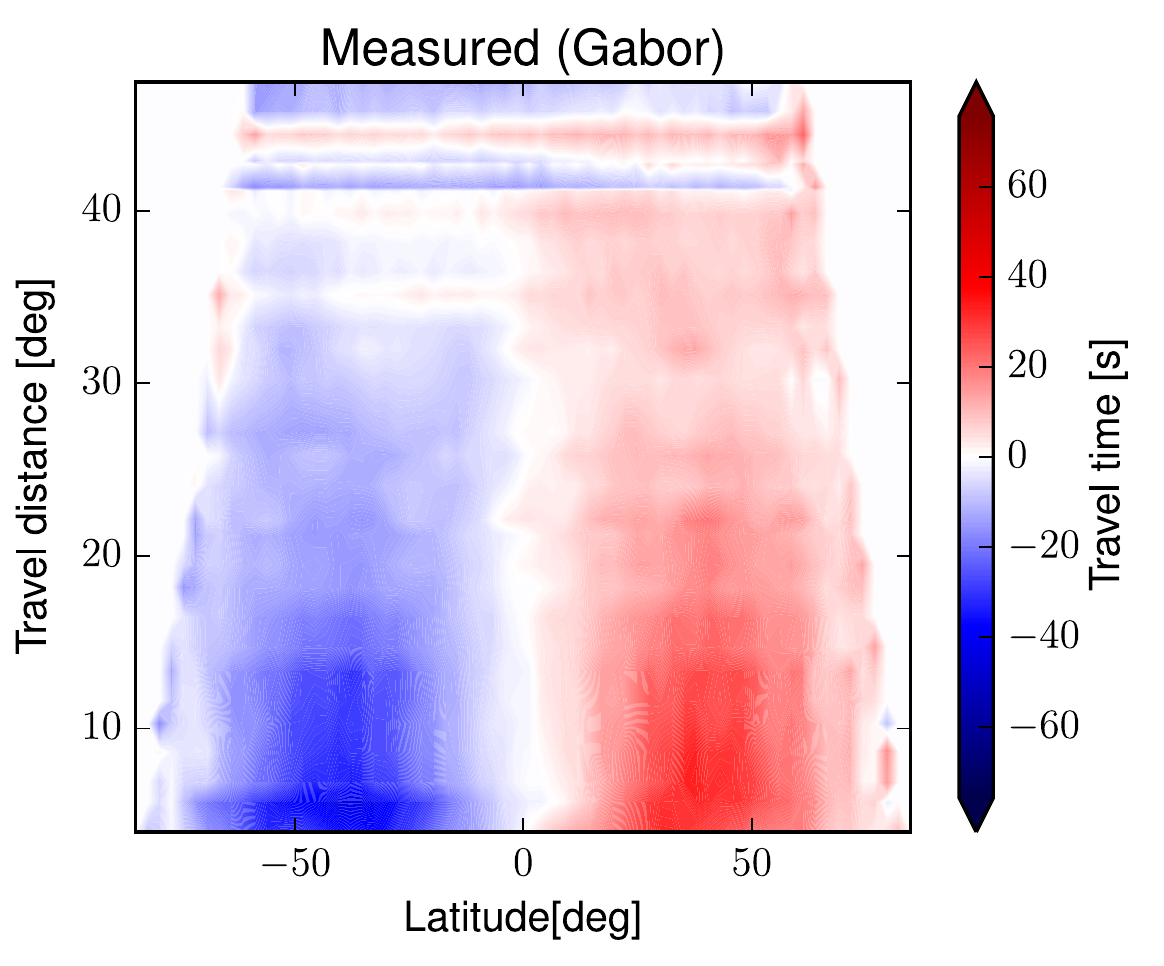}%
\includegraphics[height=1.6in]{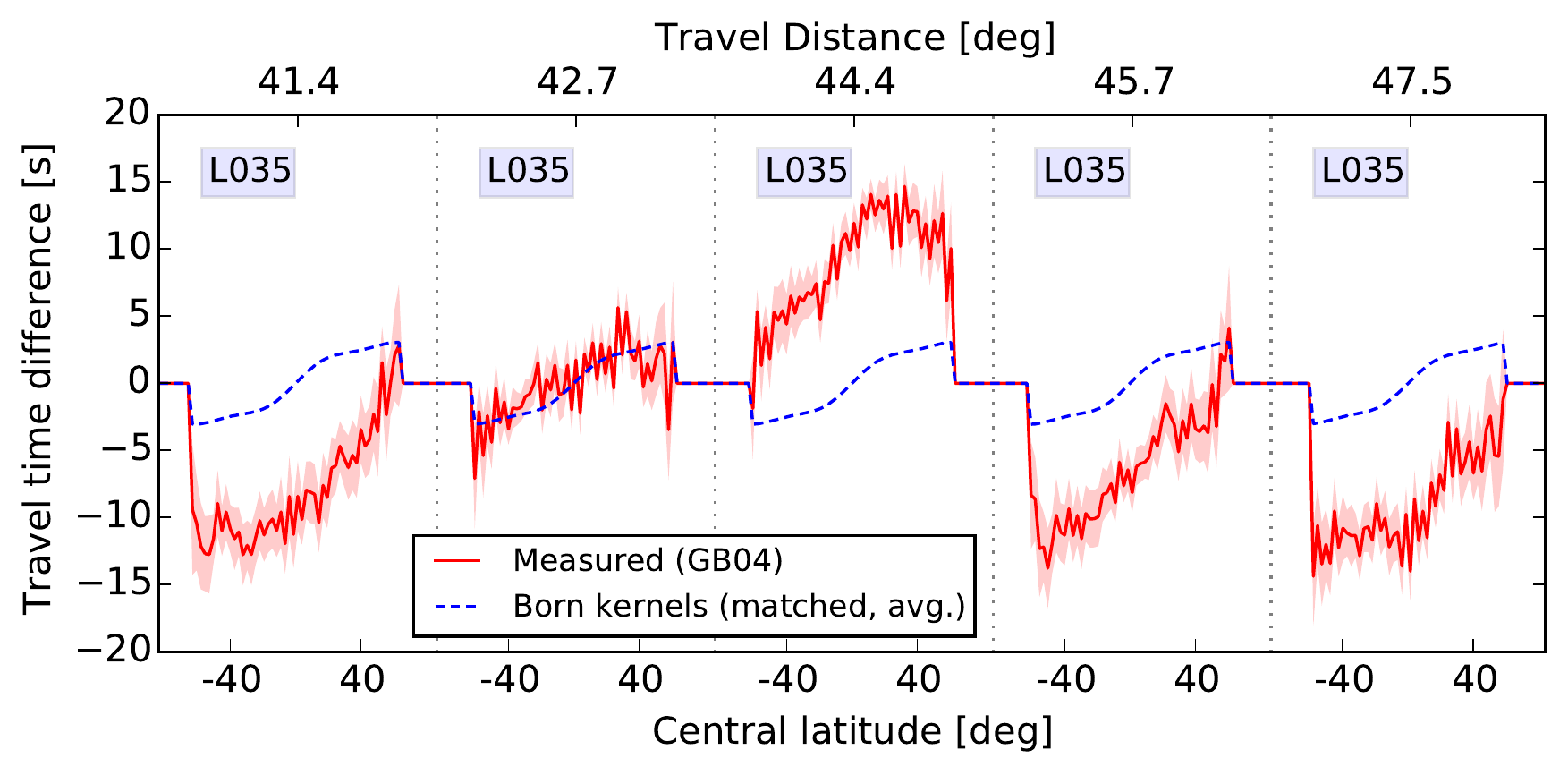}%

\caption{Measured travel times, without offset correction. Left and middle panels: same travel times as displayed in Figure~\ref{figfwdtts2d} (top left and top middle panels). Right panel: Measured GB04 and forward-modelled travel times from the Born kernels for all travel distances, at which a constant offset correction was applied in Figure~\ref{figfwdtts1d}.}%
\label{figfwdttsnooffset}%
\end{figure*}

As the influence of the radial flow component on the travel times has not been taken into account by \citet{Zhao2013} and \citet{Jackiewicz2015}, we evaluate its impact in the following using our Born approximation model, see the green dashed line in Figure~\ref{figfwdtts1d}. For small travel distances, the contribution from the radial flow to the total travel time is small and increases with travel distance. E.g., for a latitude of $40.5\degr$ and a travel distance of $\Delta=6.9\degr$, the impact of the radial flows relative to the horizontal flows is $0.3\,\%$. This ratio increases to $21.8\,\%$ for a travel distance of $\Delta=47.5\degr$. However, for all distances, the magnitude of the contribution of the radial flows is smaller than the measurement errors of the travel times.

This result is valid for the special case of the simulated data and the flow profile used in this study. Nevertheless, it indicates that it may be worthwhile to study the impact of including kernels for the radial flow component in meridional flow inversions.

\section{OUTLOOK: INVERSIONS WITH BORN KERNELS}
\label{secinversions}

\begin{figure*}%
\begin{center}

\includegraphics[height=5.5cm]{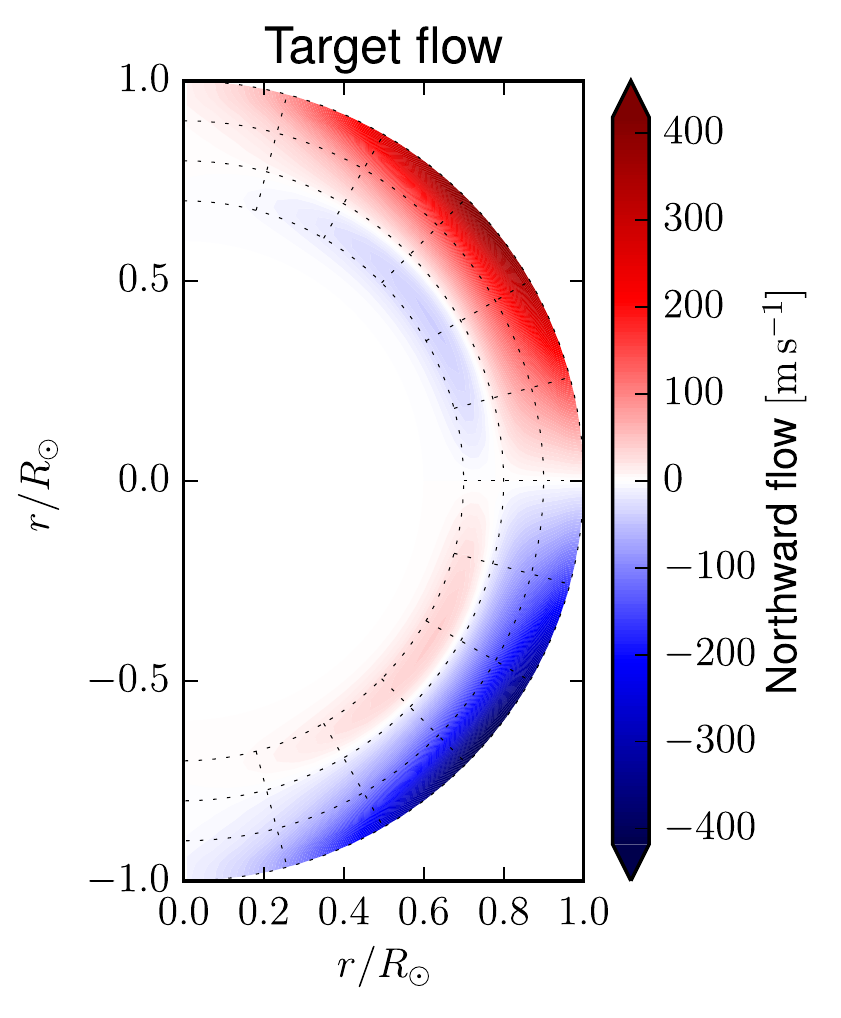}%
\includegraphics[height=5.5cm]{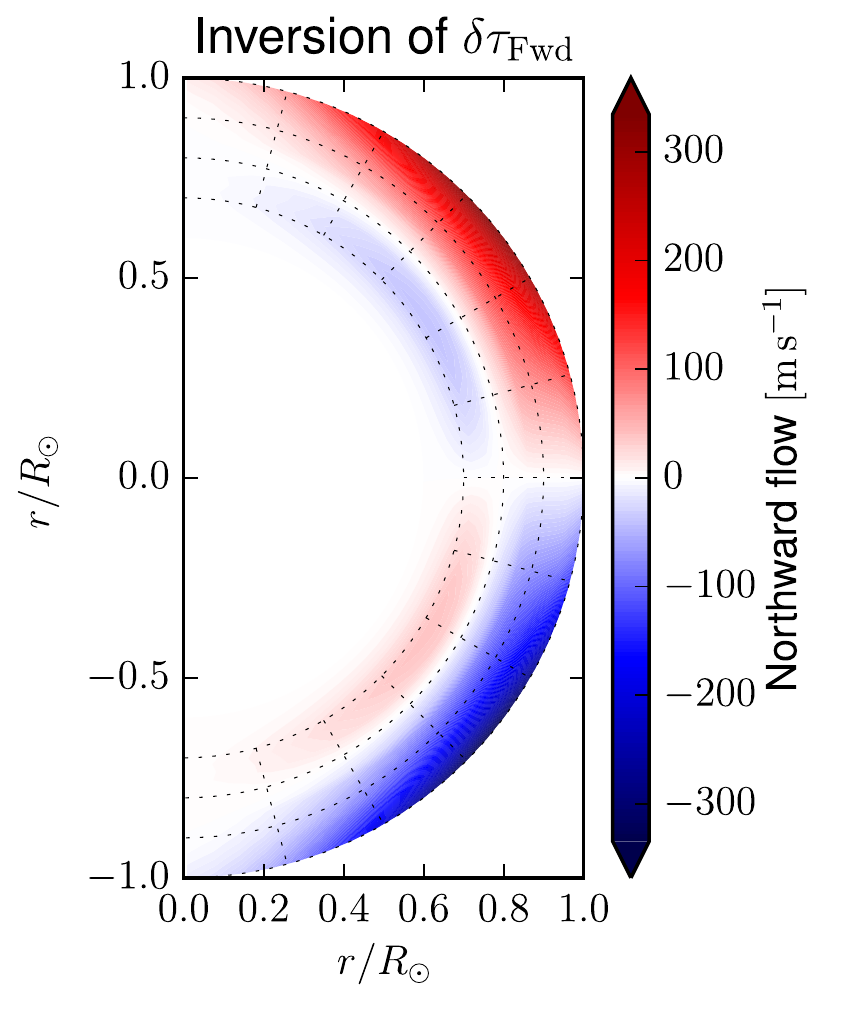}%
\includegraphics[height=5.5cm]{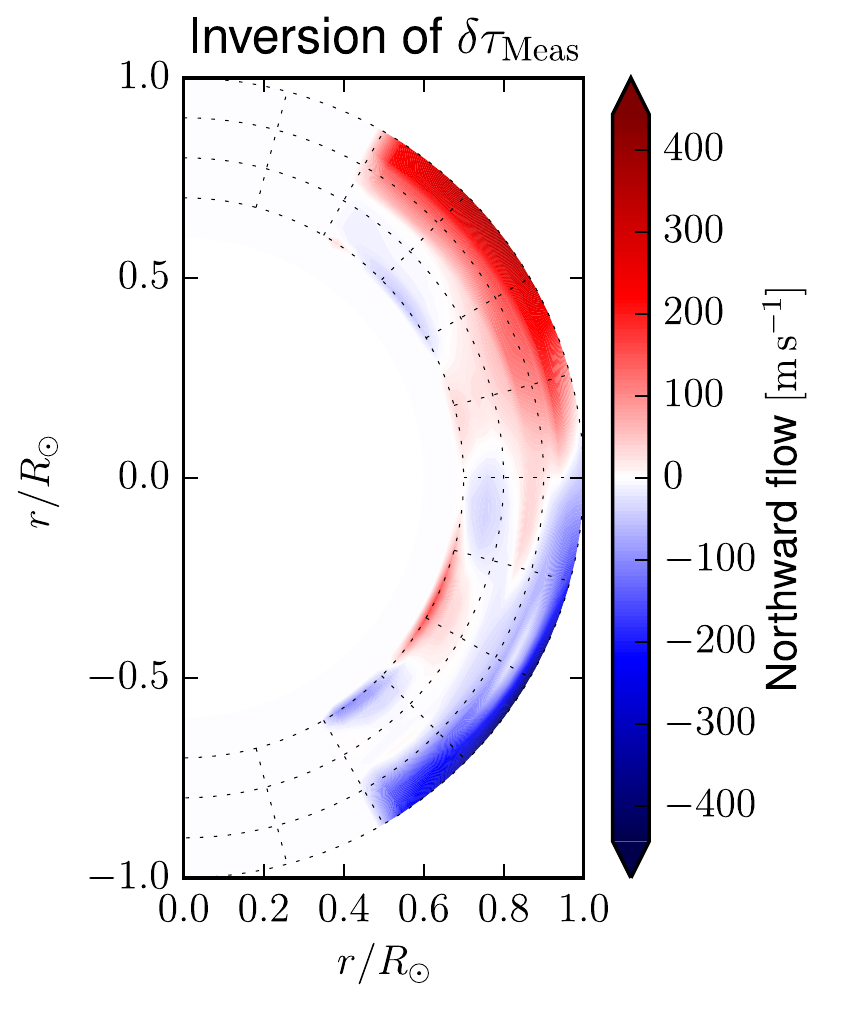}%
\includegraphics[height=5.5cm]{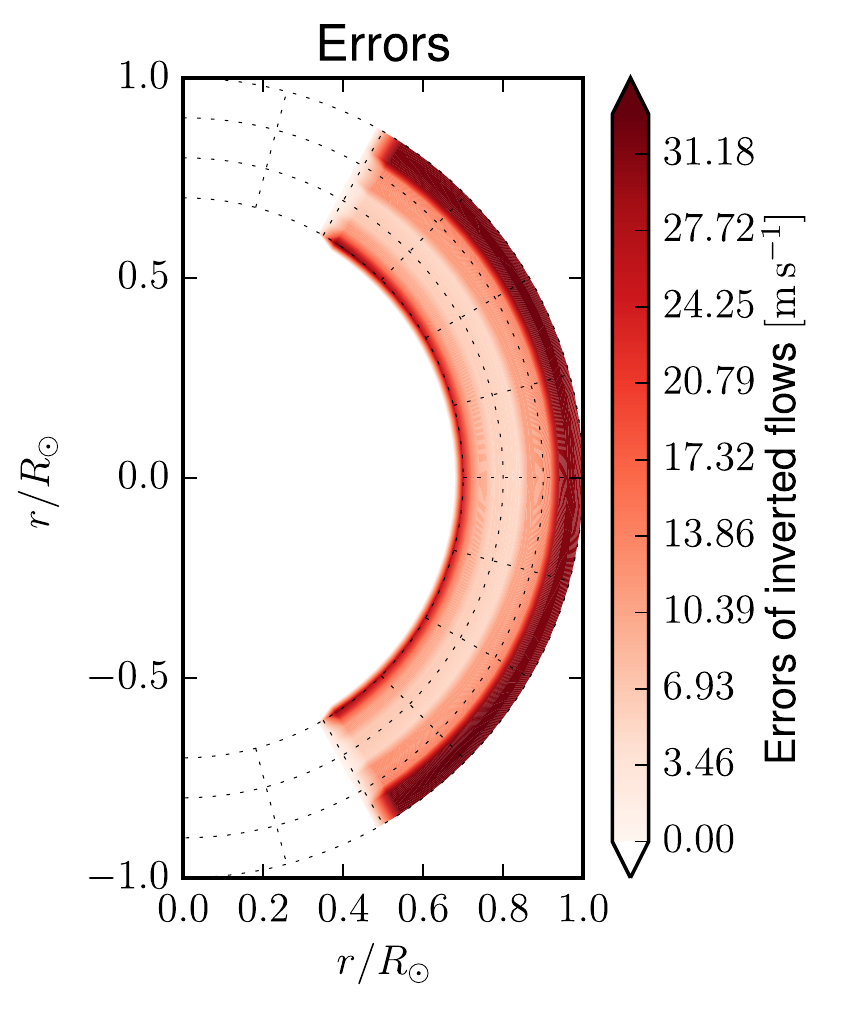}%

\end{center}

\caption{Inversion results for meridional flows. Shown are (from left to right) target flows (flow profile from the simulation convolved with target kernels), inverted flows from forward-modelled travel times ($\delta\tau_\mathrm{Fwd}$) and measured travel times ($\delta\tau_\mathrm{Meas}$), as well as inversion errors for the inversion of measured travel times. Grey dashed lines indicate locations at $r/R_\Sun = 0.7,0.8,0.9,1.0$ and multiples of 15 degrees distance from the equator.\label{figinvmeas}}%
\end{figure*}

\begin{figure*}%
\begin{center}

\includegraphics[height=5cm]{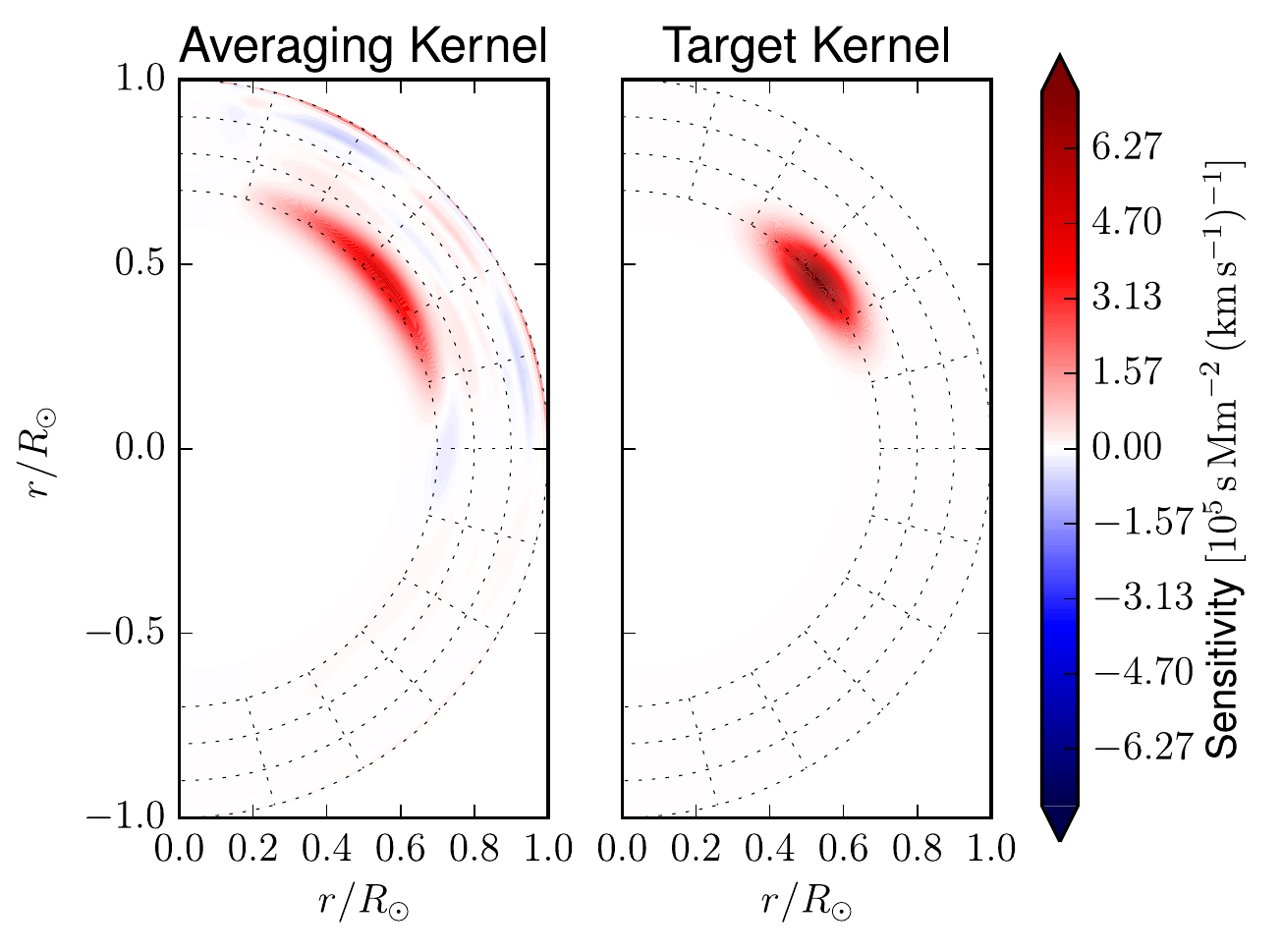}%
\includegraphics[height=5cm]{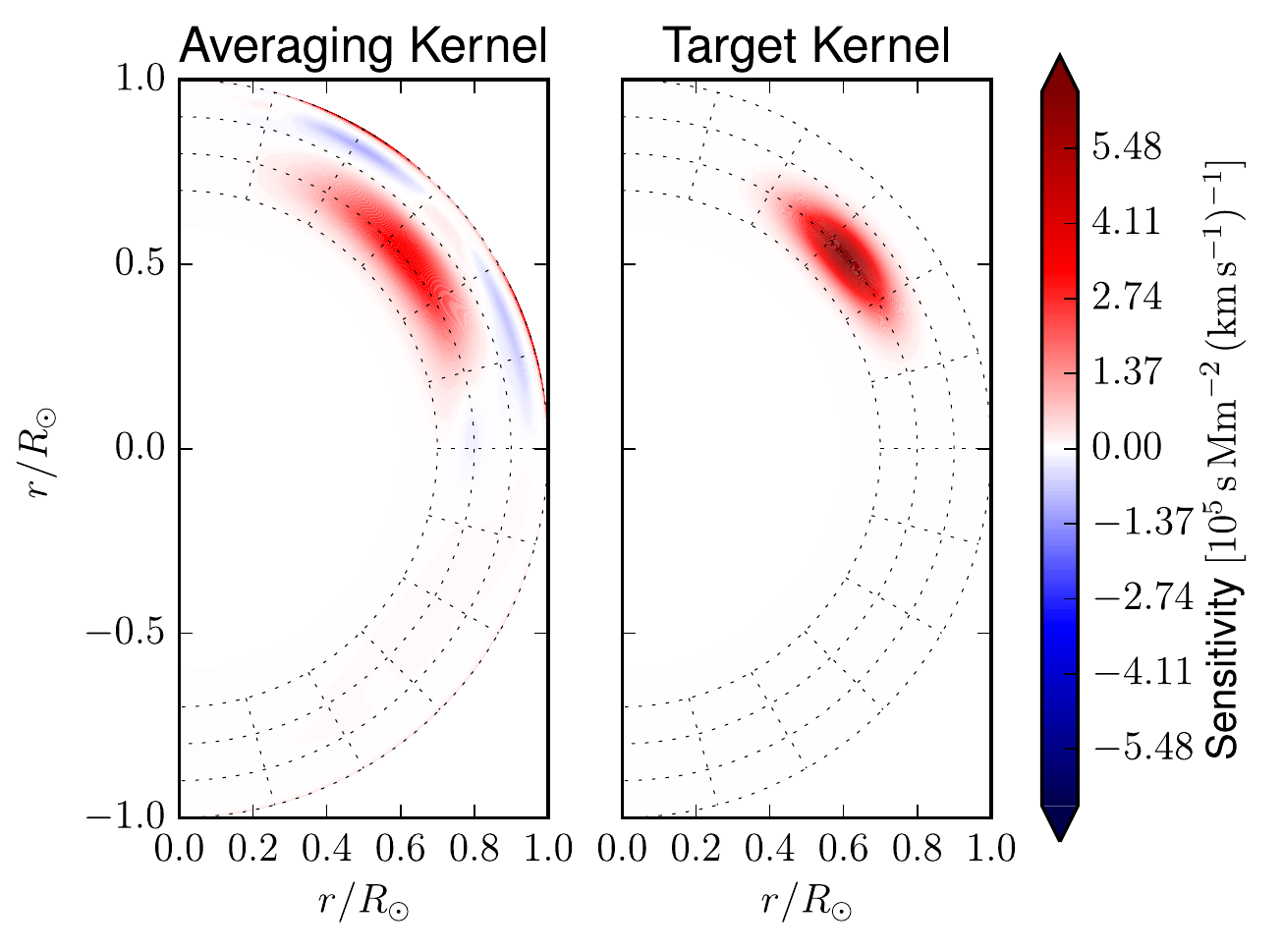}%
\includegraphics[height=5cm]{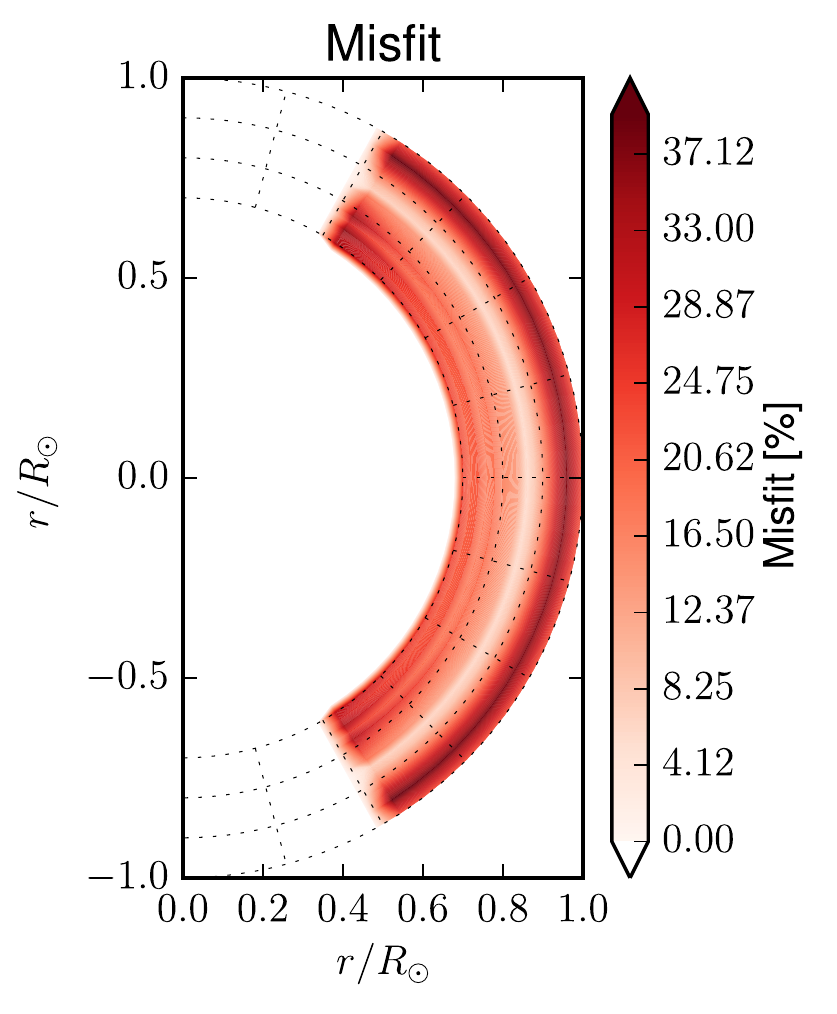}%

\end{center}

\caption{Example averaging and target kernels for a target latitude of $40\degr$ and target depths of $0.7 R_\Sun$ (left panel) and $0.8 R_\Sun$ (middle panel), as well as misfit of averaging kernels as a function of target location (right panel, see the main text for a definition). \label{figinvtargetKs}}%
\end{figure*}

In order to show that spherical Born approximation kernels can be used for inferring solar meridional flows, a standard SOLA inversion procedure for the horizontal flow component is carried out following the approach of \citet{Jackiewicz2015} using only the horizontal component of the Born kernels.

The travel times for the two smallest travel distances belonging to the lowest phase-speed filter (L170) were excluded from the inversion as they cannot be trusted, see the discussion in Section~\ref{sectts} and Figure~\ref{figfwdtts1d}.

In Figure~\ref{figinvmeas}, the inversion results are shown. The target flow profile (first panel) was obtained by convolving the target kernels with the flow profile from the simulation. The second panel shows a flow map resulting from the inversion of noiseless forward-modelled travel times. It can be seen to match qualitatively very well the target flow profile. The inversion result for the noisy measured travel times (third panel) matches the target flow in a coarser sense, recovering some but not all of the flow pattern from the simulation.

Figure~\ref{figinvtargetKs} shows averaging and target kernels for two example target locations. The misfit of the averaging kernels shown in the right panel of Figure~\ref{figinvtargetKs} was obtained with
\begin{equation}
\text{misfit}(r_\mathrm{T},\theta_\mathrm{T}) = \frac{\int (\mathcal{K} -T )^2 \,r\,\id r \, \id \theta}{\int T^2 \,r\,\id r \, \id \theta},
\label{eqmisfit}
\end{equation}
where $\mathcal{K}=\mathcal{K}(r,\theta;r_\mathrm{T},\theta_\mathrm{T})$ is the averaging kernel and $T=T(r,\theta;r_\mathrm{T},\theta_\mathrm{T})$ is the target kernel for a particular target location $(r_\mathrm{T},\theta_\mathrm{T})$. We note that the weights and thus the averaging kernels are identical for both inversions presented here.

\section{CONCLUSIONS}
\label{secconclusions}

In this paper, we have presented the validation of spherical Born kernels for inferring the deep solar meridional flow with time-distance helioseismology.

We showed that it is possible to efficiently compute spherical Born kernels for measuring the deep solar meridional flow. To do so, we used the recently developed approach of \citet{Boening2016}, which was further optimized for computational efficiency, either for obtaining two-dimensional integrated or full three-dimensional kernels. The numerical optimization was based on the horizontal variation of the eigenfunctions being separable from the radial dependence and the radial order of the mode. Compared to a recently developed method by \citet{Gizon2016}, the numerical efficiency of our method is found to be similar, with some advantages in the case of filtered kernels or in the case of a fine frequency resolution as needed for deep flow measurements.

Using a spherical Born approximation model, it is possible to accurately model observational quantities relevant for time-distance helioseismology such as the mean 
power spectrum, disc-averaged 
cross-covariances, and first-order travel times perturbed by a given flow field.

We also show that the match of the reference cross-covariance between model and observations depends on the match of the model power spectrum to the observed one. The agreement is very good if the mode frequencies and damping rates entering the model are extracted from the measured power spectrum. 
The match between observed and modelled travel times, however, does not seem to depend significantly on the match in the power spectrum for travel distances larger than $8\degr$ for the flow model considered here. For travel distances smaller than $8\degr$, we found a noticeable dependence of the forward travel times on the match of the power spectrum to observations.

Using a standard 2D SOLA inversion of travel times measured from the simulated data for the horizontal flow component, we can recover most features of the input meridional flow profile in the inverted flow map. The agreement is particularly good for the inversion of noiseless forward-modelled travel times. When inverting noisy measured travel times, we obtain a coarse agreement between inverted and target flows. This shows that Born kernels can be used for inferring the deep solar meridional flow if the noise level in the data is small enough.

The Born approximation is thus a promising method for inferring large-scale solar interior flows. We note, however, that an extensive study is needed in order to compare the use of Born and ray kernels in inversions of the deep meridional flow in more detail.

\acknowledgments

Acknowledgements: The research leading to these results has received funding from the European Research Council under the European Union's Seventh Framework Programme (FP/2007-2013) / ERC Grant Agreement
n. 307117. This work was supported by the SOLARNET project (www.solarnet-east.eu), funded by the European Commission's FP7 Capacities Programme under the Grant Agreement 312495. J.J. acknowledges support from the National Science Foundation under Grant Number 1351311. S.K. was supported by NASA’s Heliophysics Grand Challenges Research grant 13-GCR1-2-0036. The authors thank Thomas Hartlep for providing the simulated data. V.B. thanks Kolja Glogowski for computing eigenmodes for Model S and helping with many science-related IT issues. The authors acknowledge fruitful discussions during the international team meeting on “Studies of the Deep Solar Meridional Flow”  at ISSI (International Space Science Institute), Bern. The authors thank the referee for constructive comments which improved the paper.


\appendix

\section{FAST COMPUTATION OF SPHERICAL BORN SENSITIVITY FUNCTIONS}
\label{appendixoptimization}

Our starting point is from Equations (38-40) in \cite{Boening2016},
\begin{align}
        \bK(\br_1,\br_2; \br)=& \sum_{i=(\bar l,\bar n),j=(l,n)} \; J_{ij}(\br_1,\br_2) \, \bZ^{ij}(\br_1,\br_2; \br)  + \Big( 1 \leftrightarrow 2 \Big)^* \label{eqcalckernelappendix},
\end{align}
where
\begin{align}
        {\bZ^{ij}(\br_1,\br_2; \br)} &= \rho_0(r)   { \sum_{k=r,\theta,\phi} \cO_k^{\bar l \bar n} (\br)\Big[ P_{\bar l}(\cos \Delta_2)\Big]  {\bm \bnabla}_{\br} \Bigg[   \cO_k^{ln} (\br) \Big[  P_{l}(\cos \Delta_1)  \Big]     \Bigg]}  , \label{eqZij}   \\
{J_{ij}(\br_1,\br_2)} &= (2l+1)(2\bar l +1) R_{ln}(r_s) R_{\bar l \bar n}(r_2)\nonumber \\
			& \, \, \, \times { \sum_{n'}  R_{ln'}(r_s) R_{ln'}(r_1)   \int_{-\infty}^\infty  \frac{\ii \omega^3 \Wdiffstar(\br_1,\br_2,\omega)  M(\omega) \, f(l,\omega) f(\bar l,\omega)}{4\pi \, (\sigma^{2}_{ln}-\omega^2) (\sigma^{2*}_{ln'}-\omega^2)(\sigma^{2}_{\bar l \bar n}-\omega^2)} \, \id \omega  } .\label{eqJij}
\end{align}
It is now possible to separate the horizontal and radial dependence of the eigenfunctions imprinted in the operators $\cO_k^{l n}$ in Equation~\eqref{eqZij} (see \citealp{Boening2016}, Eq. (17), for their definition) so that one obtains
\begin{align}
    {\bZ^{ij}_m(\br_1,\br_2; \br)} &= \rho_0(r)    \sum_{k=r,\theta,\phi}  a_{k,i}(r) \, c_{k,\bar l}(\Omega_2,\Omega) \cdot b_{m,k,j}(r) \, d_{m,k,l}(\Omega_1,\Omega), \label{eqZabcd}
\end{align}
using appropriate definitions for $a,b,c,d$. From Equations \eqref{eqZabcd} and \eqref{eqcalckernel} we derive, bearing in mind $i=(\bar l,\bar n),j=(l,n)$,
\begin{align}
    {\bm K}_m(r,\theta,\phi) 
					&=  \rho_0(r)  \, \sum_{\bar l, l} \, \sum_{k=r,\theta,\phi} \, c_{k,\bar l}(\Omega_2,\Omega) \, d_{m,k,l}(\Omega_1,\Omega) \, T_{m,k,\bar l,l}(r;\br_1,\br_2)     + \Big( {\bf 1 \leftrightarrow 2 }\Big)^* \\
				&=  \rho_0(r)  \, \sum_{\bar l, l} \, \sum_{k=r,\theta,\phi} \, q_{m,k,\bar l,l}(\Omega_1,\Omega_2,\Omega) \, T_{m,k,\bar l,l}(r;\br_1,\br_2)     + \Big( {\bf 1 \leftrightarrow 2 }\Big)^*  \label{eqcalckernelsuperfastappendix}
\end{align}
where we defined
\begin{align}
    T_{m,k,\bar l,l}(r;\br_1,\br_2)  &= \sum_{\bar n, n} \; J_{(\bar n, \bar l),(n,l)}(\br_1,\br_2)  \; a_{k,(\bar n, \bar l)}(r) \, b_{m,k,(n, l)}(r). \label{eqT}
\end{align}


%

\end{document}